\documentclass[Journals,InsideFigs,NoLineNumbers]{ascelike}

\usepackage{graphicx}
\usepackage{multirow} 
\usepackage[centertags]{amsmath} \usepackage{amsfonts}
\usepackage{amsthm}
\usepackage{upgreek}
\usepackage{dblfloatfix}
\usepackage{cleveref}
\DeclareMathAlphabet{\mathpzc}{OT1}{pzc}{m}{it}
\usepackage{amsmath}
\usepackage{widetext}
\usepackage{flushend}
\usepackage{afterpage}
\usepackage{cuted}
\usepackage{bbm}
\usepackage[mathscr]{euscript}
\usepackage{gensymb}
\usepackage{color}
\usepackage{url}
\usepackage{times}
\usepackage[utf8]{inputenc}
\usepackage[T1]{fontenc}
\usepackage{lmodern}
\usepackage{newtxtext,newtxmath}
\usepackage[noblocks]{authblk}
\usepackage{natbib}
\usepackage[figurename=Fig.,labelfont=bf,labelsep=period]{caption}

\begin{document}

\title{Generalized Stratified Sampling for Efficient Reliability Assessment of Structures Against Natural Hazards}

\author{Srinivasan Arunachalam, S.M.ASCE}
\affil{Graduate Student, Department of Civil and Environmental Engineering, University of Michigan, Ann Arbor, MI 48109. E-mail: sriarun@umich.edu}
\author{Seymour M.J. Spence, Ph.D., A.M.ASCE}
\affil{Associate Professor, Department of Civil and Environmental Engineering, University of Michigan, Ann Arbor, MI 48109 (corresponding author). E-mail: smjs@umich.edu}

\maketitle
\begin{abstract}
Performance-based engineering for natural hazards facilitates the design and appraisal of structures with rigorous evaluation of their uncertain structural behavior under potentially extreme stochastic loads expressed in terms of failure probabilities against stated criteria. As a result, efficient stochastic simulation schemes are central to computational frameworks that aim to estimate failure probabilities associated with multiple limit states using limited sample sets. In this work, a generalized stratified sampling scheme is proposed in which two phases of sampling are involved: the first is devoted to the generation of strata-wise samples and the estimation of strata probabilities whereas the second phase aims at the estimation of strata-wise failure probabilities. Phase-I sampling enables the selection of a generalized stratification variable (i.e., not necessarily belonging to the input set of random variables) for which the probability distribution is not known \textit{a priori}. To improve the efficiency, Markov Chain Monte Carlo Phase-I sampling is proposed when Monte Carlo simulation is deemed infeasible and optimal Phase-II sampling is implemented based on user-specified target coefficients of variation for the limit states of interest. The expressions for these coefficients are derived with due regard to the sample correlations induced by the Markov chains and the uncertainty in the estimated strata probabilities. The proposed stochastic simulation scheme reaps the benefits of near-optimal stratified sampling for a broader choice of stratification variables in high-dimensional reliability problems with a mechanism to approximately control the accuracy of the estimators of multiple failure probabilities. The practicality and efficiency of the scheme are demonstrated using two examples involving the estimation of failure probabilities associated with highly nonlinear responses induced by wind and seismic excitations. 
\end{abstract}


\section{Introduction}
\label{Sec:Intro}

The advancement of computing power, algorithms, and frameworks in the last couple of decades has enabled the analysis of engineering systems with greater scrutiny than ever before. However, computational models are not perfect simulators of real-world systems/behaviour, and the real world itself is uncertain. Uncertainty in model and parameter selection can be characterized using random variables, processes, fields and waves capturing both epistemic uncertainties (arising from the lack of knowledge/data) as well as aleatory uncertainties (arising from intrinsic randomness of phenomena) \citep{shinozuka1991simulation,shinozuka1996fields,gurley1997analysis,gurley1999applications,der2009aleatory,melchers2018structural}. For practical problems, the effect of the input uncertainty on the model outputs is of prime importance to characterize safety against violation of multiple constraints (or limit states) through failure probabilities, or equivalently, reliabilities. The determination of the failure probability of a component, or a system, $P_{f,h}$ involves solving the following $N_d$-dimensional integral:
\begin{equation}
    \label{ProbInt}
    P_{f,h} = P(\pmb{\Theta} \in \Gamma_h) = \int_{\mathbb{S}}{\mathbbm{1}_{f,h}(\pmb{\uptheta})}q(\pmb{\uptheta})d\pmb{\uptheta}
\end{equation}
where $\pmb{\uptheta} \in \mathbb{S} \subset \mathbb{R}^{N_d}$ is a realization of the $N_d$-dimensional vector of basic random variables $\boldsymbol{\Theta}$ with joint probability density function (PDF) $q$; $\Gamma_h$ is the failure region within the sample space $\mathbb{S}$ associated with the $h$th limit state; $\mathbbm{1}_{f,h}(\pmb{\uptheta})$ is an indicator function assuming a value of 1 if $\pmb{\uptheta} \in \Gamma_h$ and 0 otherwise. For most applications in natural hazards engineering, the following characteristics make the estimation of $P_{f,h}$ of Eq. \eqref{ProbInt} challenging: (i) a high-dimensional uncertain space, $N_d$ in the order of several thousands, necessary for accommodating the white noise sequence, $\boldsymbol{\Theta_Z}$, modeling load stochasticity; (ii) the need to simultaneously evaluate multiple nonlinear limit state functions (LSFs), $\mathcal{G}_h$ with $h = \{1,2,\ldots,H\}$ where $H$ is the total number of associated performance objectives; and (iii) the need to estimate small failure probabilities (e.g., $P_{f,h} \leq 10^{-4}$) at affordable computational costs while maintaining acceptable accuracy for engineering applications.

The outcrossing method is one of the widely used methods to treat time-variant reliability problems. However, the outcrossing rate can be analytically calculated only for several special cases where a set of strong assumptions can be made about the responses (and their derivatives) for the application of the (generalized) Rice formula \citep{rice1944mathematical,derkiureghian_2022,li2022explicit}. In general, the limit state functions can be characterized by strongly non-Gaussian and non-stationary responses. Approximate calculations of the outcrossing rate may also be difficult in light of the above challenges associated with Eq. \eqref{ProbInt}. Monte Carlo (MC) methods are the simplest of simulation-based uncertainty quantification techniques and are robust to the dimension of the uncertainties as well as the number and nature of the limit states. However, they suffer from the need to carry out a large number of system evaluations, $n$, if small failure probabilities are to be estimated with sufficient accuracy (e.g., $n = 10^{k+2}$ samples are required to estimate a $P_{f,h}$ in the range of $10^{-k}$ with a $10\%$ coefficient of variation). This is often computationally prohibitive for complex computational models with significant nonlinearities, and/or with fine discretization in space/time. A vast literature exits on variance reduction techniques for reducing the computational burden associated with MC simulation. Importance sampling modifies the sampling density function so as to draw more samples from the ``important region'' of $\mathbb{S}$ \citep{melchers1989importance,fishman2013monte}. However, identifying the optimal importance sampling density (ISD) is generally difficult and when the choice of the form of ISD adopted is inappropriate, the variability of the estimator cannot be controlled in the presence of a large number of uncertain parameters \citep{au2003importance}. Importance sampling and its variants (e.g., \cite{au1999adaptive,papaioannou2016sequentialIS}), as well as other methods, such as line sampling \citep{koutsourelakis2004linesampling,schueller2004critical} and subset simulation (SuS) \citep{au2001originalsubset,au2003subsetseismic}, are based on generating samples that better probe the failure region such that a larger proportion of them contribute to the evaluation of the failure probabilities. SuS is based on the idea of estimating small failure probabilities as a product of larger conditional probabilities by introducing intermediate failure events. Although the original algorithm \citep{au2001originalsubset} focuses on evaluating the failure probability of a single rare failure event (i.e., associated with a single LSF), some variants have been proposed that generalize the approach to multiple LSFs \citep{hsu2010PSubSim,li2015GSS,li2017systemGSS2}. In contrast, the class of simulation schemes based on stratified designs includes, but is not limited to, stratified random sampling \citep{cochran2007sampling}, Latin Hypercube Sampling (LHS) \citep{stein1987large}, and Partially Stratified Sampling (PSS) \citep{shields2016}. These represent better sampling plans owing to improved space-filling properties but may not be particularly focused on any failure region, or LSF. Surrogate-assisted approximation techniques aim to replace the expensive simulator (the LSF or the limit state surface) with an emulator (e.g., polynomial chaos expansion, kriging surrogates \citep{sudretmetamodel}) built from a so-called design of experiments, a set of observed points to approximate the true function/surface. However, they are usually unsuitable for high-dimensional and highly-nonlinear problems. 

Conventional stratified sampling is limited to applications where efficient stratification can be defined by directly specifying intervals for the components of $\boldsymbol{\Theta}$ with known joint PDF, $q$. It is not generally applicable to a wider set of problems in which a potential efficient stratification variable can be identified as the output of an auxiliary computational model or the output of an intermediate computational model belonging to the model chain used to estimate the system response. This can be a significant limitation when solving reliability problems in performance-based engineering for natural hazards that pose the following challenges: (i) response quantities, defining the LSFs of interest, that generally require the evaluation of a cascade of computational models for characterizing the hazard, hazard-structure interaction, structural response, and loss/damage; (ii) LSFs for which intervals directly defined on a subset of $\boldsymbol{\Theta}$ do not represent an efficient stratification; (iii) indicator functions characterizing the exceedance of LSFs of interest that are expensive to evaluate due to the need to evaluate a cascade of high-fidelity models; (iv) performance targets involving small failure probabilities, or equivalently, large reliabilities. The post-stratification technique is rarely useful since the probability distribution of variables outside of $\boldsymbol{\Theta}$ is rarely available. In a similar formulation to that of stratified sampling, the double sampling procedure requires two phases of sampling; the first is devoted to the generation of strata-wise samples and the estimation of strata probabilities whereas the second phase aims at the estimation of strata-wise failure probabilities. In this paper, an extended double-sampling-based stochastic simulation scheme is proposed to estimate multiple failure probabilities for a suite of limit states with a built-in optimization procedure to control the estimation errors while using limited samples sets. The novelty in the proposed scheme is the Markov Chain Monte Carlo (MCMC)-driven Phase-I sampling similar to subset simulation when MC simulation is deemed infeasible and the optimal execution of Phase-II sampling based on user-specified target coefficients of variation (c.o.v) for the limit states of interest. The expressions for these coefficients are derived with due regard to the sample correlations induced by the Markov chains and the uncertainty in the estimated strata probabilities. The proposed scheme is illustrated using two examples involving the estimation of failure probabilities associated with highly nonlinear responses induced by wind and seismic excitations. 


\section{Background}
\label{Sec:Background}

The basic idea of stratified sampling is to define partitions of the sample space, $\mathbb{S}$, such that samples are drawn from each of these partitions (or strata), $\{\mathbb{S}_i: i = 1,\ldots,m\}$, in a preferred manner. This implies that the user can decide the stratification variables, denoted by the vector $\pmb{\upchi}$, the strata boundaries as well as the number of samples within each stratum, $n_i$. The strata need to satisfy: $\cup_{i=1}^{m}\mathbb{S}_i = \mathbb{S}$ and $\mathbb{S}_i \cap \mathbb{S}_j = \emptyset$ for $i\neq j$. As a result, Eq. \eqref{ProbInt} can be broken down into sub-integrals as:
\begin{equation}
   \label{IntSubInt}
    P_{f,h} = \sum_{i=1}^m \int_{\mathbb{S}_i}{\mathbbm{1}_{f,h}(\pmb{\uptheta})}q(\pmb{\uptheta})d\pmb{\uptheta}
\end{equation}
Since for the conditional PDF the following holds: $q(\pmb{\uptheta} \mid \mathbb{S}_i) = q(\pmb{\uptheta}){\mathbbm{1}_{\mathbb{S}_i}(\pmb{\uptheta})}/ P(\mathbb{S}_i)$; $P_{f,h}$ can be further simplified as:
\begin{equation}
   \label{IntSubInt2}
    P_{f,h} = \sum_{i=1}^m \int_{\mathbb{S}_i}{\mathbbm{1}_{f,h}(\pmb{\uptheta})}q(\pmb{\uptheta}\mid\mathbb{S}_i)P(\mathbb{S}_i)d\pmb{\uptheta} = \sum_{i=1}^m P_{f_i,h}P(\mathbb{S}_i)
\end{equation}
where $P(\mathbb{S}_i)=$ the volume of the $i$th stratum in the probability space and $P_{f_i,h}=$ the conditional failure probability. When MC sampling is performed within each stratum, the procedure is known as stratified random sampling and $P_{f,h}$ is approximated as:
\begin{equation}
   \label{Estimatorexp}
    P_{f,h} \approx \tilde{P}_{f,h} = \sum_{i=1}^m {\sum_{k=1}^{n_i}{\mathbbm{1}_{f,h}(\pmb{\uptheta}^{(i)}_k)}P(\mathbb{S}_i)}/n_i
\end{equation}
where ${\pmb{\uptheta}^{(i)}_k}=$ the $k$th independent and identically distributed (i.i.d) sample out of $n_i$ samples in the $i$th stratum. Clearly, the decomposition of the integral of Eq. \eqref{ProbInt} is enabled by the theorem of total probability. In particular, $\tilde{P}_{f,h}$ of Eq. \eqref{Estimatorexp} can be seen as a weighted sum of $\tilde{P}_{f_i,h}$ with the weights, $P(\mathbb{S}_i)$. More importantly, $P(\mathbb{S}_i)$ is perfectly known only when stratification is directly performed by specifying lower and upper bounds for each component of $\pmb{\upchi}$, with $\pmb{\upchi} \subseteq \boldsymbol{\Theta}$, since, under these circumstances, $q(\pmb{\upchi})$ is available. Moreover, the simulation of i.i.d samples, ${\pmb{\uptheta}^{(i)}_k}$, is straightforward as the conditional density, $q(\pmb{\uptheta}\mid \mathbb{S}_i)$, can be obtained from the joint density $q(\pmb{\uptheta})$. The variance reduction achieved through stratified random sampling is dependent on the choice of $\pmb{\upchi}, \{\mathbb{S}_i\}_{1\leq i \leq m}$ and $\{n_i\}_{1\leq i \leq m}$. A poor implementation could potentially lead to a worse performance than direct MC simulation.

Stratified sampling was developed in the survey sampling community, wherein stratification based on demographic features is commonly employed 
for estimation of sub-population characteristics/parameters \citep{cochran2007sampling,arnab2017surveybook}. The incorporation of the exact probability weights (i.e., stratum probabilities) corrects for differences in the distribution of the traits/features in the sample set and in the actual population which explains the unconditional variance reduction when proportional sample allocation (i.e., $n_i = nP(\mathbb{S}_i)$) is considered. In some instances, when a fixed number of samples cannot be generated from each stratum due to the choice of $\pmb{\upchi}$, classification of samples into their respective strata can be performed after sampling, a procedure termed post-stratification. Post-stratification assumes that the strata probabilities are known accurately and only that the stratum to which a sample belongs is unknown 
\citep{cochran2007sampling,GLASGOW2005}. Further, when even the strata probabilities are not known \textit{a priori}, a large simple random sample set can be drawn to first estimate the strata probabilities and prepare a pool of samples for each stratum from which a smaller sample set can be used to evaluate the failure probabilities. This technique is known as double sampling since the process involves a first phase of sampling devoted to strata construction, strata-wise sample classification, and estimation of strata probabilities before carrying out a second phase of sampling for estimating the failure probabilities of interest through stratification \citep{cochran2007sampling,GLASGOW2005,Rao1973double}.
This paper focuses on the development of a generalized stratified sampling scheme for risk assessment problems in natural hazards engineering through adoption of double sampling methods. Specifically, improving the computational efficiency in double sampling (i.e., $\pmb{\upchi} \nsubseteq \boldsymbol{\Theta}$) through both optimal sample allocation as well as adoption of Markov MCMC to accelerate Phase-I sampling is investigated.

\section{Proposed Double-Sampling-based Simulation Scheme}
\subsection{Simulation of Strata-wise Samples}
\subsubsection{Basic idea of double sampling}

As discussed earlier, if $\pmb{\upchi} \subseteq \boldsymbol{\Theta}$ and $q(\pmb{\uptheta}\mid \mathbb{S}_i)$ is known, the generation of strata-wise input samples is trivially achieved by sampling $q(\pmb{\uptheta}\mid \mathbb{S}_i)$ through MC simulation, a task that generally requires minimal computational effort. Consider now $\pmb{\upchi} = \mathcal{H}(\pmb{\upsigma})$ with $\mathcal{H}$ a computational model that depends on a subset of the input uncertainties, $\pmb{\upsigma}$, with the remaining input uncertainties (assumed to be independent of $\pmb{\upsigma}$ for simplicity) denoted with $\pmb{\uptau}$ so that $\pmb{\uptheta} = \{\pmb{\upsigma}, \pmb{\uptau}\}$. For example, if peak hourly-mean wind speed is selected as the stratification variable and it is not a basic random variable, then $\mathcal{H}$ can denote the function mapping (i.e., the hazard model) between the random wind hazard parameters (constituting $\pmb{\upsigma}$) and the wind speed (i.e., $\pmb{\upchi}$). The remainder of the uncertainties, say, concerning system uncertainties and uncertainties in aerodynamics will constitute $\pmb{\uptau}$. Clearly, the choice of the stratification variable defines the computational model $\mathcal{H}$ for the problem. If for a given problem, the cost of evaluating $\mathcal{H}$, denoted as $\mathscr{C}(\mathcal{H})$, is much less relative to the cost of evaluating the limit state functions, $\mathscr{C}(\mathcal{G}_h) \forall h$, then a MC simulation can be adopted to generate a large number of samples such that the requisite number of samples in every stratum, $\{n_i\}_{1\leq i \leq m}$ is available. It should be observed that while this does produce i.i.d samples ${\pmb{\uptheta}^{(i)}_k}$ in each stratum, if $P(\mathbb{S}_m) \approx 10^{-k}$, then it takes $10^{k+2}$ evaluations of $\mathcal{H}$ to generate roughly $10^2$ samples in $\mathbb{S}_m$, i.e., the last stratum, which will yield an estimate of $P(\mathbb{S}_m)$ with a c.o.v of $10\%$. In particular, the estimator is given by the expression:
\begin{equation}
    \label{condsample_MC}
    \begin{split}
        \tilde{P}_{f,h} & = \sum_{i=1}^m \tilde{P}_{f_i,h} \tilde{P}(\mathbb{S}_i) \\ & =
        \sum_{i=1}^m \left(\frac{\sum_{k=1}^{n_i} {\mathbbm{1}_{f,h}(\pmb{\uptheta}^{(i)}_k)}}{n_i}\right )
        \frac{\hat{n_i}}{\hat{n}}
    \end{split}
\end{equation}
${\pmb{\uptheta}^{(i)}_k} = [{\pmb{\upsigma}^{(i)}_k},\pmb{\tau}_k]$ where $\pmb{\tau}_k=$ non-conditional MC samples; $\hat{n}=$ the total number of MC samples generated out of which $\hat{n_i}$ lie in the $i$th stratum; while $n_i \le \hat{n_i}$ are the samples utilized in the calculation of conditional failure probabilities. This implies that $n = \sum_i n_i$ limit state evaluations are performed in total, whereas $\hat{n} = \sum_i \hat{n}_i$ evaluations of $\mathcal{H}$ are performed to populate samples within strata and to estimate the stratum probabilities. It is noted that in the literature, the consideration of $\pmb{\uptau}$ and its separate MC sampling has not been explicitly described but is essential to this work. 

An important property of the classic stratified sampling of Section ``Background'' is the utilization of the knowledge of accurate probability weights which is lost here. Its implications can be observed as follows: (i) if $\hat{n}_i = n_i$, then Eq. \eqref{condsample_MC} reduces to simple MC estimation of ${P}_{f,h}$. Therefore, it is required that $\hat{n} >> n$ such that $\tilde{P}(\mathbb{S}_i)$ is a relatively high-accuracy estimate, which is feasible since $\mathcal{H}$ is cheap to evaluate; (ii) proportional sample allocation (i.e., $n_i = nP(\mathbb{S}_i)$), which guarantees variance reduction for classic stratified sampling regardless of $\pmb{\upchi}$ and $\{\mathbb{S}_i\}_{1\leq i \leq m}$, loses this guarantee since it again reduces the scheme to simple MC estimation. This emphasizes how for high efficiency gains, the sample allocation needs to mirror, as much as possible, the theoretical optimal allocation, a problem that is discussed in Section ``Sample Allocation Scheme''. Let $\tilde{\tilde{P}}_{f_i,h}$ define  the estimate of ${P_{f_i,h}}$ when $n_i = \hat{n}_i$, then the variance can be written as (Theorem 1, \cite{Rao1973double}):
\begin{equation}
    \label{Varexp2}
    \begin{split}
        \mathbb{V}({\tilde{P}_{f,h}}) & =  \mathbb{V}\left(\sum_{i=1}^m {\tilde{P}_{f_i,h}} \tilde{P}(\mathbb{S}_i)\right) \\& = \mathbb{V}\left(\sum_{i=1}^m \tilde{\tilde{P}}_{f_i,h} \tilde{P}(\mathbb{S}_i) + \sum_{i=1}^m {(\tilde{P}_{f_i,h} -\tilde{\tilde{P}}_{f_i,h})} \tilde{P}(\mathbb{S}_i)\right)\\& =
        \mathbb{V}\left(\frac{\sum_{k=1}^{\hat{n}} {\mathbbm{1}_{f,h}(\pmb{\uptheta}^{(i)}_k)}}{\hat{n}} \right) + \mathbb{V}\left(\sum_{i=1}^m {(\tilde{P}_{f_i,h} -\tilde{\tilde{P}}_{f_i,h})} \tilde{P}(\mathbb{S}_i)\right)
        \\& =
        \mathbb{V}\left(\frac{\sum_{k=1}^{\hat{n}} {\mathbbm{1}_{f,h}(\pmb{\uptheta}^{(i)}_k)}}{\hat{n}} \right)\\&+
        \mathbb{E}\left(\mathbb{V}\left(\sum_{i=1}^m {(\tilde{P}_{f_i,h} -\tilde{\tilde{P}}_{f_i,h})} | \tilde{P}(\mathbb{S}_i) \right) \tilde{P}(\mathbb{S}_i)\right) 
        \\& =
        \frac{P_{f,h}(1-P_{f,h})}{\hat{n}} + \sum_{i=1}^m {\frac{P(\mathbb{S}_i)P_{f_i,h}(1-P_{f_i,h})}{\hat{n}}\left(\frac{1}{\nu_i} -1\right)}        
    \end{split}
\end{equation}
where $\mathbb{E}=$ the expectation operator, $\nu_i = n_i/\hat{n}_i \in (0,1] = $ the sub-sampling fraction whose value is assumed to be fixed and which represents the proportion of samples in the $i$th stratum from Phase-I considered in Phase-II for failure probability evaluations. In the above derivation, the following results were used \citep{Rao1973double,cochran2007sampling}: $\text{Cov}(\tilde{\tilde{P}}_{f_i,h},\tilde{P}_{f_i,h} -\tilde{\tilde{P}}_{f_i,h}) = 0$, $\mathbb{E}(\tilde{P}_{f_i,h}) = \tilde{\tilde{P}}_{f_i,h}$, and $\mathbb{V}(\tilde{P}_{f_i,h} -\tilde{\tilde{P}}_{f_i,h}) = \mathbb{V} (\tilde{P}_{f_i,h}) - \mathbb{V}(\tilde{\tilde{P}}_{f_i,h})$. Notably, the first summand of the final expression of Eq. \eqref{Varexp2} is fixed for a given limit state and $\hat{n}$, whereas the second summand represents the sample-allocation-dependent variance contribution which vanishes as $n_i \to \hat{n}_i$. The estimator is unbiased and consistent in the sense that it approaches the true failure probability as $\hat{n} \to \infty$, for fixed $\nu_i$. Finally, the c.o.v can be estimated as:
\begin{equation}
    \label{covMC}
    \begin{split}
        \kappa_h = \frac{\sqrt{\mathbb{V}({\tilde{P}_{f,h}})}}{P_{f_i,h}} \approx \frac{
        \sqrt{\frac{\tilde{P}_{f,h}(1-\tilde{P}_{f,h})}{\hat{n}} + \sum_{i=1}^m {\frac{\tilde{P}(\mathbb{S}_i)\tilde{P}_{f_i,h}(1-\tilde{P}_{f_i,h})}{\hat{n}}\left(\frac{1}{\nu_i} -1\right)}}
        }{\sum_{i=1}^m \tilde{P}_{f_i,h} \tilde{P}(\mathbb{S}_i)}
    \end{split}
\end{equation}
%


\subsubsection{Extension through subset simulation for high-efficiency gains}

It is inefficient to use MC simulation when $\mathscr{C}(\mathcal{H})$ is not trivial and, in particular, when $P(\mathbb{S}_m)$ is extremely small. The latter might be necessary when rare subspaces of $\pmb{\upchi}$ (lying in the tail of its joint PDF) are of special interest in producing extreme responses. In such cases, a more efficient technique is required to populate strata-wise samples and approximate strata probabilities. The class of methods based on MCMC algorithms can achieve adaptive sample generation from conditional distributions (conditional on $\mathbb{S}_i$) \citep{papaioannou2015mcmc}. For instance, sequential importance sampling can be applied to produce conditional samples by a transition of samples through a sequential reweighting operation whose governing distribution sequence gradually approaches the target conditional distribution \citep{papaioannou2016sequentialIS}. In this paper, owing to its wider usage, SuS is considered for efficient Phase-I sampling \citep{au2001originalsubset}. Unlike the traditional application of SuS, in this work, SuS only provides sufficient samples in each stratum to enable a stratified sampling-based estimation of multiple failure probabilities.  

Consider a single stratification variable denoted by $\chi \in [\chi_L,\chi_U]$, then by fixing the thresholds $\chi_i$, where $\chi_0 < \chi_1 < \ldots < \chi_{m-1} < \chi_m$, the strata, $\{\mathbb{S}_i\}_{1\leq i \leq m}$, and nested intermediate event sequence, $F_1 \supset F_2 \supset \ldots \supset F_{m-1}$ are defined as follows: $F_i = \{\pmb{\uptheta}: \chi>\chi_i\}, \forall i \leq (m-1)$ and $\mathbb{S}_i = \{\pmb{\uptheta}: \chi \in (\chi_{i-1},\chi_i]\}, \forall i \leq m$. It is also notationally convenient to define $F_0 = \mathbb{S}$, a certain event. The last stratum, $\mathbb{S}_m = F_{m-1}$, is bounded from above by $\chi_m = \chi_U$ (which need not be finite) and from below by $\chi_0 = \chi_L$ to ensure the satisfaction of the probability partition properties. The adaptive procedure of SuS generates samples in $F_{i}$ (and $\mathbb{S}_{i+1}$) by simulating states of Markov chains through MCMC starting from the samples (or seeds) conditional on $F_{i-1}, \forall i \leq (m-1)$ \citep{au2001originalsubset,papaioannou2015mcmc}. It can be proved that for an idealized version of the SuS method with fixed thresholds, the optimal choice of thresholds is to make the conditional probabilities $P(F_i|F_{i-1})$ equal \citep{bect2017BSS}. This provides the rationale for the widely adopted idea of fixing the sample estimate of $P(F_i|F_{i-1}),  \forall i \leq (m-1)$ to be $p \in [0.1,0.3]$, a constant such that $\chi_i$ and $\mathbb{S}_i$ are adaptively defined. In other words, $\chi_i$ is chosen as the $(1-p)$th quantile of the conditional samples in $F_{i-1}$. It is easy to note that $\tilde{P}(F_i) = p^i$ and $\tilde{P}(\mathbb{S}_i) = p^{i-1}(1-p), \forall i \leq (m-1)$, where tilde denotes that the quantity is a sample estimate. Let the total number of Markov chain samples in each conditional level of $F_i$ be $N$, then the number of Markov chain samples generated in the $i$th stratum for $\forall i \leq (m-1)$ will be , $\hat{n}_i = (1-p)N$ with $\hat{n}_m = N$, from which it follows that $\hat{n} = N(m(1-p)+p)$. The values of $n_i$, however, are determined according to the optimal allocation scheme of Section ``Sample Allocation Scheme''. Both within each stratum and among strata, the generated samples, ${\pmb{\uptheta}^{(i)}_k} = [{\pmb{\upsigma}^{(i)}_k},\pmb{\tau}_k]$ are correlated through $\pmb{\upsigma}^{(i)}_k$ due to inherent correlation of the Markov chains, while $\pmb{\tau}_k$ are uncorrelated as they are i.i.d MC samples unaffected by the SuS, or stratification procedures. The variance expressions need to take into account both the sample correlations induced by SuS as well as the uncertainty in the estimated strata probabilities. Appendix \ref{appdixA} discusses the properties of $\tilde{P}_{f_i,h}$, $\tilde{P}(\mathbb{S}_i)$, and $\tilde{P}_{f,h}$. This includes the derivation of the variance of $\tilde{P}_{f,h}$ that enables the introduction of the following expression for the estimator c.o.v. of the extended scheme:
 \begin{equation}
    \label{covSuS}
     \kappa_h \approx \frac{\sqrt{\sum_{i=1}^m \tilde{\vartheta}_{i,h}^2 \left(\tilde{\vartheta}_{\mathbb{S}_i}^2 + \tilde{P}^2(\mathbb{S}_i) \right) + \sum_{i=1}^m \sum_{j=1}^m \tilde{P}_{f_i,h} \tilde{P}_{f_j,h} \tilde{\vartheta}_{\mathbb{S}_{ij}}^2}}{\sum_{i=1}^m \tilde{P}_{f_i,h} \tilde{P}(\mathbb{S}_i)}
\end{equation}   
where $\tilde{\vartheta}_{i,h}^2=$ the estimate of $\mathbb{V}(\tilde{P}_{f_i,h})$, $\tilde{\vartheta}_{\mathbb{S}_{ij}}^2=$ the estimate of $\text{Cov}(\tilde{P}(\mathbb{S}_i),\tilde{P}(\mathbb{S}_j))$, and $\tilde{\vartheta}_{\mathbb{S}_i}^2=$ the estimate of $\mathbb{V}(\tilde{P}(\mathbb{S}_i))$, all of which can be estimated using the simulated Markov chain samples and evaluation of the limit state violations. Notably, the estimates, $\tilde{\vartheta}_{\mathbb{S}_i}^2$ and $\tilde{\vartheta}_{\mathbb{S}_{ij}}^2$ are dependent only on the Phase-I samples, and independent of the limit states and Phase-II sampling. On the other hand, the estimate $\tilde{\vartheta}_{i,h}^2$ is dependent on the Phase-I samples, $n_i$, and the $h$th limit state function. This implies that for a given problem, the variance component $\sum_{i=1}^m \sum_{j=1}^m \tilde{P}_{f_i,h} \tilde{P}_{f_j,h} \tilde{\vartheta}_{\mathbb{S}_{ij}}^2$ of Eq. \eqref{covSuS} is independent of the sample allocation (i.e., of $\{n_i\}_{1\leq i \leq m}$) and only reflects the adequacy of Phase-I sampling. 

\begin{figure}
\centering
\includegraphics[width = \textwidth]{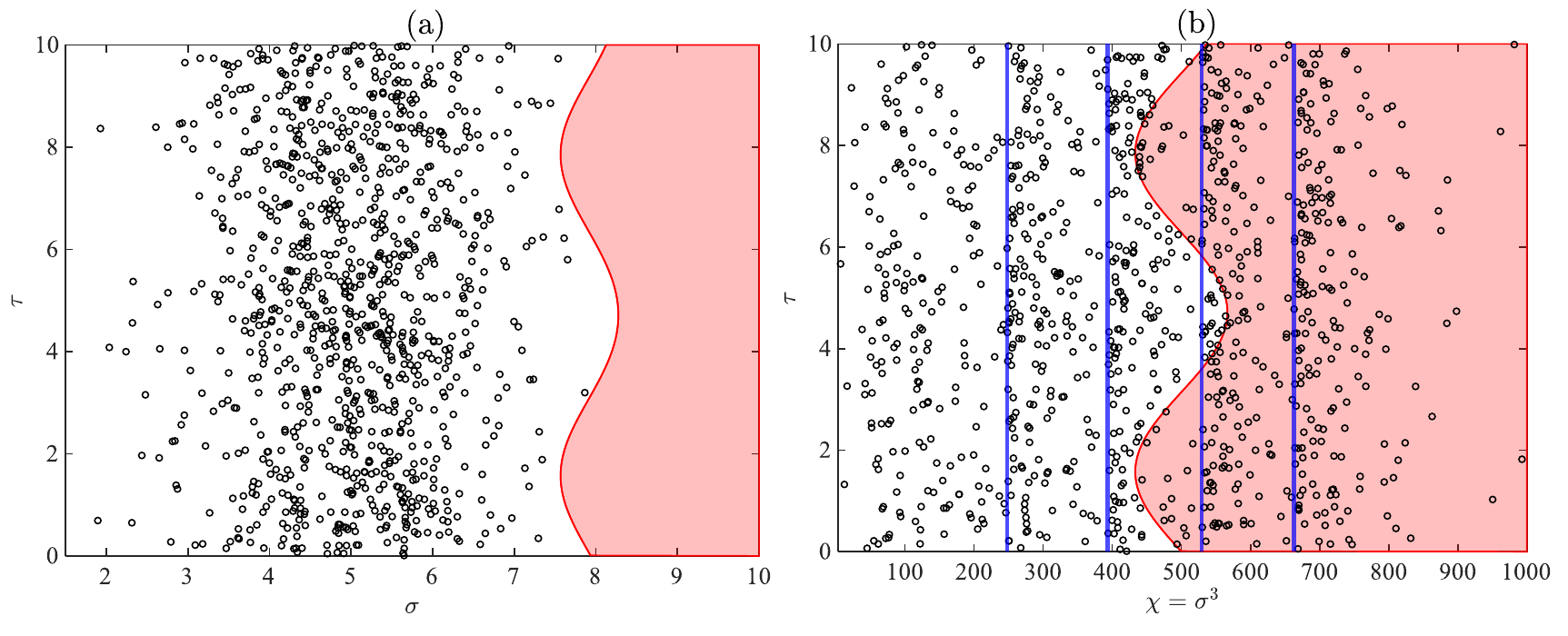}
\caption{Illustration of sample distribution in (a) MC simulation; (b) proposed simulation scheme.}
\label{fig:sampledistribution_illus}
\end{figure}

For a conceptual illustration of the proposed method, consider a two-dimensional problem with two independent random variables $\tau \sim U(0,10)$ and $\sigma \sim N(5,1)$. Figure \ref{fig:sampledistribution_illus}(a) illustrates the distribution of 1000 samples in a Monte Carlo simulation for estimating the failure probability associated with the failure region (red-shaded in the figure) given by $200\sin(\tau) + 3\sigma^3 > 1500$. In contrast, if $\chi = \mathcal{H}(\sigma) = \sigma^3$ is desired to be the stratification variable, then the proposed method generates strata-wise samples to enable strata-wise calculation of the failure probabilities. Figure \ref{fig:sampledistribution_illus}(b) illustrates how the proposed method efficiently explores the failure region, wherein the vertical solid lines indicate the strata separation (according to $p = 0.1$ and $m = 5$), and $\tau$ is randomly sampled from its marginal distribution. For the simplicity of this illustration, equal sample allocation (i.e., 200 samples in each stratum) was adopted. Further, the application of subset simulation has not been demonstrated since $\chi$ is simply $\sigma^3$, however, it should be noted that, in general, $\chi$ is a complex function of two or more random variables, in which case subset simulation becomes necessary to generate the strata-wise samples.

\subsubsection{Additional remarks}

For the more general case of multiple stratification variables, the same framework can be realized by replacing the SuS algorithm with the generalized subset simulation (GSS) algorithm, originally developed as an extension of SuS for estimating multiple failure probabilities using a single run of the simulation scheme \citep{li2015GSS,li2017systemGSS2}. Basically, in the aforementioned SuS procedure, $\{F_i\}$ are determined using a single driving variable, $\chi$, whereas in GSS unified intermediate events (i.e., $F_i = \{\pmb{\uptheta}: \chi^{(1)}>\chi^{(1)}_i\} \cup \{\pmb{\uptheta}: \chi^{(2)}>\chi^{(2)}_i\} $ for two stratification variables $\chi^{(1)}$ and $\chi^{(2)}$) can be defined to drive samples to multiple strata. However, this modification can be cumbersome in providing sufficient samples in all strata and does not lend itself to calculable variance expressions that are required for the optimal sample allocation procedure, central to the proposed simulation scheme. 

It is worth mentioning that while the development of this extension was independent, it bears some similarities with the parallel subset simulation (P-SuS) algorithm \citep{hsu2010PSubSim} and the response conditioning method (RCM) proposed by \cite{au2007augmenting}. The key idea in P-SuS is to introduce a principal variable that is correlated with all performance functions, as the driving variable in SuS, and multiple failure probabilities are estimated simultaneously. Here, the principal variable is a representative output variable (e.g., an average of the maximum story drifts) such that each simulation will not only provide a realization of the principal variable but also of all performance functions (e.g., the maximum story drifts for all stories) at once without requiring any additional simulation/computation. This can be seen as a special case of the proposed scheme wherein Phase-II sampling/simulation (including the uncertainties given by $\pmb{\tau}$) is absent. On the other hand, RCM leverages information from computationally inexpensive approximate solutions to the target problem to achieve efficient and consistent reliability estimates. The ``conditioning response'' which approximates the target response is stratified and SuS enables the conditional sample generation. However, the method was not directed toward reliability problems with multiple limit states, and neither of the two methods optimally evaluate samples from each stratum which is indeed actualized in this paper through a constrained-optimization-based sample allocation procedure. Further, in contrast to subset simulation where parametrization of the failure domain is necessary, the proposed method is agreeable to a more generic limit state representation, such as structural collapse, for which a non-binary measurable limit state function cannot always be assigned.

\subsection{Choice of Stratification Variables}

The gains from stratification can be significant if the choice of $\pmb{\upchi}$ is such that the stratification defined by $\{\mathbb{S}_i\}_{1 \leq i \leq m}$ promotes more intra-stratum homogeneity (with respect to the $h$th limit state violation) than the overall homogeneity in $\mathbb{S}$. The intra-stratum homogeneity can be measured by the \textit{unit variance} of the MC conditional probability estimator (i.e., associated with one simple random sample) given by ${P}_{f_i,h}(1-{P}_{f_i,h})$. In fact, the ideal stratification variable for $P_{f,h}$ is the $h$th limit state function, $\mathcal{G}_h$, itself. Obviously, it is not possible to stratify according to decreasing values of a limit state function and therefore justifying the adoption of one or more variables for stratification that are highly correlated with the response(s) of interest. Additionally, in the proposed scheme, since the stratification is based on random variables and is independent of the limit states, the same sample set within each stratum can be used to estimate the strata-wise failure probabilities for all limit states. That is, it is not necessary to rerun the simulation for each limit state of interest. By broadening the scope of selection (i.e., $\pmb{\upchi} \nsubseteq \boldsymbol{\Theta}$), a good candidate for $\pmb{\upchi}$ can be selected from the output of any intermediate model (from the sequence of numerical models that is typically involved in response estimation) or from the output of an auxiliary model not used in the modal chain. However, every choice is associated with a corresponding computational effort, proportional to $\mathscr{C}(\mathcal{H})$, to simulate strata-wise samples. In natural hazard applications, by leveraging expert knowledge, or physical intuition, good candidates for $\pmb{\upchi}$ can take the form of hazard intensity measures such as maximum hourly wind speed, the geometric mean of spectral accelerations, or the elastic base moments of wind excited systems. In general, when explicit hazard modeling is involved in a natural hazard application, the intensity measure is itself an output of a numerical model and the probability distribution may not be typically known. Further, the proposed scheme enables the consideration of hybrid stratification variables such as a certain combination of the peak wind speed and mean hourly rainfall intensity for an application where both the wind and the concurrent rainfall fields could crucially affect the performance of a building system (e.g., cladding performance). It should be emphasized that the proposed estimator is unbiased and consistent (i.e., convergent to the true probability with increasing computational effort, that is for $N \to \infty$ and $n_i \to \infty$) as shown in Appendix \ref{sec:A3}.

Stratified sampling suffers from the ``curse of dimensionality'' since full stratification in $k$ dimensions with $m$ strata per dimension quickly causes an explosion in the number of strata, $m^k$, and the sampling demands to meet certain accuracy in the \textit{unit variance} estimation needed for optimal sample allocation, and consequently, the estimated failure probabilities \citep{PHARR2017747}. This encourages thoughtful selection of one or two variables for stratification that strongly affect the responses, which is usually not difficult to identify from the intermediate model inputs/outputs in natural hazard applications. The number of strata, $m$ is typically determined by $P(\mathbb{S}_m)$ and the order of the smallest probability, $\min_h P_{f_i,h}$, however, increasing $m$ beyond 10 will seldom be profitable as it increases the sampling demands, or contributes to increased estimator variance arising from large uncertainty in the \textit{unit variance} estimations and sub-optimality of sample allocation for fixed sampling costs \citep{cochran2007sampling}.

\subsection{Sample Allocation Scheme}
\label{Sec:sampleallocation}
In addition to the choice of $\pmb{\upchi}$ and $\{\mathbb{S}_i\}_{1 \leq i \leq m}$, the allocation of samples among the strata defined by $\{n_i\}_{1 \leq i \leq m}$ affects the variance reduction for a fixed number of limit state evaluations, $n$. For a single limit function, the optimal allocation, termed ``Neyman allocation'', assigns samples to strata in proportion to $P(\mathbb{S}_i)$ as well as the square root of the \textit{unit variance} \citep{neyman1934,cochran2007sampling,esrel21}. For multiple LSFs, since any sample allocation cannot be simultaneously Neyman optimal for all LSFs, the solution to the following c.o.v-based constrained optimization problem needs to be considered:
    \begin{equation}
    \label{OptEqn}
    \begin{aligned}
    &\min_{\{n_i\}_{1\leq i \leq m}} &\qquad& n = {\sum_{i=1}^m n_i} &\\
    &\text{subject to:}               &      & \kappa_h \left( n_1,\ldots,n_m \right) \leq \omega_h &\quad h\leq H\\
    &                                &      & n_i \leq \hat{n}_i                                   &\quad i \leq m
    \end{aligned}
    \end{equation}
where $\kappa_h \left( n_1,\ldots,n_m \right)=$ the sample-allocation-dependent c.o.v of $\tilde{P}_{f,h}$ whereas $\omega_h=$ the user-specified c.o.v target for controlling the estimation accuracy. The ``optimal solution'' to the above-formulated problem is denoted as $\{ \Breve{n}_i \}_{1 \leq i \leq m}$ and can be found using any gradient-based optimization technique. However, the c.o.v calculation requires the knowledge of the \textit{unit variances} for all limit state functions and strata, the unavailability of which requires one to conduct a preliminary study \citep{evans1951stratification}. The goal of the preliminary simulation-based study, say using $n_p$ samples in each stratum, is purely to enable the resolution of Eq. \eqref{OptEqn} (for efficiently allocating the remaining $n-n_p$ samples) by quantifying the intra-stratum variability associated with the estimated failure probabilities associated with the selected LSFs. The preliminary study can be viewed as an exploration step carried out prior to the exploitation step of optimally executing Phase-II sampling to estimate the failure probabilities. It is important to mention that the preliminary study may introduce a systematic error in estimation, referred to as cardinal error, associated with misrepresenting any of the \textit{unit variances} as zero due to inadequate exploration \citep{amelin2004KTH,esrel21}. This can be avoided to some extent through careful strata construction and by imposing a constant lower limit on $n_i$ $ \forall i$.

\begin{figure}[]  
   \centering
    \includegraphics[height=0.9\textheight]{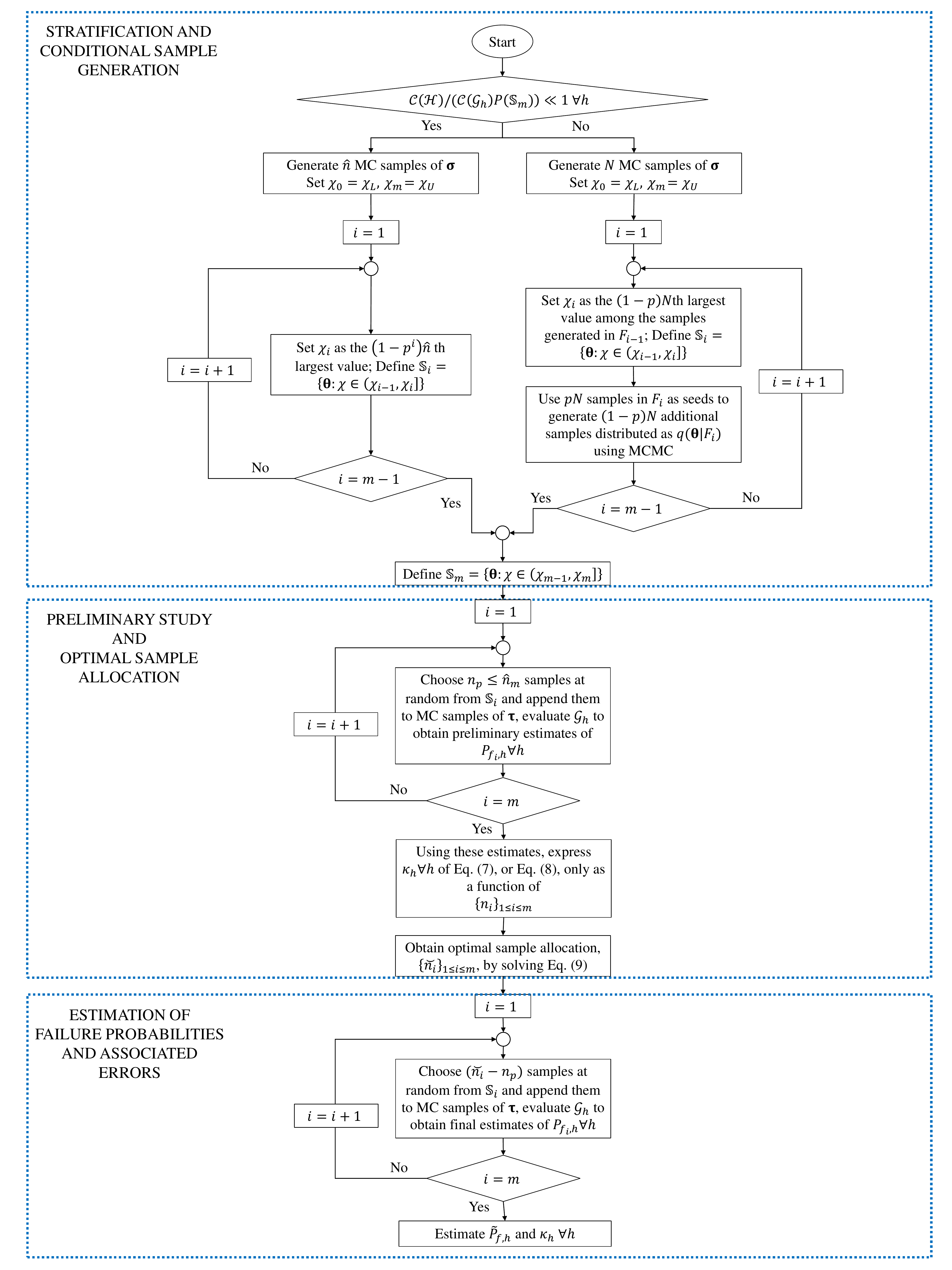}
    \caption{Flowchart of the proposed stochastic simulation procedure.}
    \label{fig:algo}
\end{figure}
 \subsection{Overall Algorithm}
 
The proposed procedure is summarized as follows:
\begin{enumerate}
    \item \textit{Initialization}: Choose a stratification variable, $\chi$, the number of strata, $m$, and probability constant, $p \in [0.1,0.3]$, defining the stratification and fixing the estimates of the strata probabilities. 
    \item \textit{Stratification and conditional sample generation}: If $\mathscr{C}(\mathcal{H})/\left(P(\mathbb{S}_m)\mathscr{C}(\mathcal{G}_h) \right) \ll 1, \forall h$, then a MC-based Phase-I sampling is feasible, else consider a subset-simulation-based sampling. If MC-based Phase-I sampling is adopted, select the total number of Phase-I samples $\hat{n}$, or, if SuS-based Phase-I sampling is adopted, select $N$. Choose the number of preliminary test samples in each stratum, $n_p$. Populate strata-wise samples, ${\pmb{\upsigma}^{(i)}_k}$, and define $\chi_i$ and $\mathbb{S}_i$ adaptively in the process.
    \item \textit{Preliminary study and optimal sample allocation}: Conduct preliminary study using $n_p$ samples drawn at random from each stratum (along with MC samples of $\pmb{\tau}$) to obtain first-level estimates of the failure probabilities with which Eq. \eqref{OptEqn} is solved to obtain $\{ \Breve{n}_i \}_{1 \leq i \leq m}$. If SuS-based Phase-I sampling is adopted, the calculation of $\kappa_h$ involves sample estimates of $\vartheta_{\mathbb{S}_i}^2$, $\vartheta_{\mathbb{S}_{ij}}^2$, $\vartheta_{i,h}^2$, and $P_{f_i,h}$.
    \item \textit{Estimation of failure probabilities and associated errors}: Using $\Breve{n}_i$ samples in $\mathbb{S}_i$, the conditional failure probabilities are estimated, combined with the strata probabilities to estimate the overall failure probabilities and their associated c.o.vs using either Eq. \eqref{covMC} or Eq. \eqref{covSuS}.
\end{enumerate}

When the preliminary-study-based optimal sample allocation roughly matches the true optimum, it is expected that the c.o.v $\kappa_h$ will be close to the respective targets, $\omega_h$, while only utilizing limited computational effort. The proposed procedure is summarized in the flowchart of Figure \ref{fig:algo}.

 The scheme can also be used in a sub-optimal form if equal sample allocation is adopted. Such an application will avoid the need to perform a preliminary study followed by optimal sample allocation. Further, if measures of accuracy in the final estimates are not required, then the implementation of the scheme will require no calculations of Eq. \eqref{covMC} or Eq. \eqref{covSuS}.


\section{Case Study}
\subsection{Example 1: Wind-excited 45-story RC building}
\subsubsection{Overview}

A 45-story reinforced concrete (RC) building of height, $H$ = 180.6 m, story height, $h_s$ = 4 m, subjected to extreme wind loads is considered to illustrate the simultaneous estimation of exceedance probabilities using the proposed methodology. The structure is assumed to be located in New York City, and the hazard model is based on the simulation of full hurricane tracks characterized by the combination of a storm track model \citep{vickery1995prediction}, wind field model \citep{jakobsen2004comparison} and a filling-rate model \citep{vickery1995wind}. The evolving wind velocity field is modeled at the site of the building through time-varying hourly mean wind speed at the building height, $v_H(t)$, and time-varying direction, $\alpha(t)$, to which a fully non-stationary and non-straight stochastic wind load model is calibrated \citep{ouyang2021performance}. In this example, peak hourly-mean wind speed, $\hat{v}_H = \max_t v_H(t)$ is chosen as the stratification variable as it is highly correlated with the responses of interest, yet is itself an output of the hurricane hazard model and therefore appropriate for the demonstration of the presented scheme. The following six responses of interest define the limit state functions: peak roof drift ratio in two orthogonal directions, $\hat{\Upsilon}_{\text{X,roof}}$ and $\hat{\Upsilon}_{\text{Y,roof}}$; residual inter-story drift ratio (IDR), $\Upsilon^{(r)}_{\text{X}}$ and $\Upsilon^{(r)}_{\text{Y}}$, and finally peak IDR over the building height, $\hat{\Upsilon}_{\text{X}}$ and $\hat{\Upsilon}_{\text{Y}}$. Two thresholds are considered for the peak roof drift ratio: 1/400, associated with the operational performance objective \citep{Prestandard19} and 1/200, associated with the continuous occupancy performance objective \citep{Prestandard19}. A threshold of 1/1000 is selected for the residual IDRs corresponding to the continuous occupancy objective \citep{Prestandard19} and 1/200 for the peak IDRs. The consideration of peak roof drifts in the reliability assessment is to limit sway at the building top and avoid issues with elevator operation/alignment whereas the consideration of residual IDRs is to limit permanent deformation due to inelastic responses \citep{Prestandard19}. The peak and residual IDRs in each orthogonal direction are reported as absolute values at the story location where the largest values occur. It can be noted that the results of the structural analyses within each stratum permit the evaluation of all the limit state functions at once and the calculation of the strata-wise failure probabilities.

\subsubsection{Stochastic wind loads}

Description of the full evolution of a hurricane event is realized through a parametric hurricane model that simulates hurricane tracks as straight lines crossing a circular sub-region centered at the building site. The outputs $v_H(t)$ and $\alpha(t)$ are modeled as functions of the distance between the building site and the eye of the hurricane, along with the consideration of the pressure decay following landfall \citep{vickery1995prediction,vickery2000simulation,vickery1995wind,jakobsen2004comparison,ouyang2021performance}. The stratification variable, $\hat{v}_H$ is dependent on the hurricane track input parameters, $\boldsymbol{\Phi}$, composed of the initial central pressure difference, $\Delta p_0$, translation speed, $c$, size of the hurricane, $r_M$, approach angle, $\theta_{\text{app}}$, minimum distance, $d_{\text{min}}$, between the building site and the hurricane track, and the coefficients $a_0$, $a_1$, and $\epsilon_f$ of the filling-rate model. Consequently, the mean annual rate of exceeding a given wind speed, $\lambda_{\hat{v}_H}$, also known as the non-directional hurricane hazard curve, can be expressed as:
\begin{equation}
    \label{windhazardcurve}
    \lambda_{\hat{v}_H} (v') = \lambda_{\text{hurr}} \int_{v'}^{\infty} \left( \int_{\boldsymbol{\Phi}} f_{\hat{v}_H | \boldsymbol{\Phi}} (v|\boldsymbol{\Phi}) f_{\boldsymbol{\Phi}} (\boldsymbol{\Phi}) d\boldsymbol{\Phi} \right) dv
\end{equation}
where $f_{\hat{v}_H | \boldsymbol{\Phi}}=$ the PDF of $\hat{v}_H$ conditional on $\boldsymbol{\Phi}$, $f_{\boldsymbol{\Phi}}=$ the joint PDF of the components of $\boldsymbol{\Phi}$, and $\lambda_{\text{hurr}} = 0.67$ is the mean annual recurrence rate of the site-specific hurricanes. The expression in parenthesis of Eq. \eqref{windhazardcurve} is equal to $f_{\hat{v}_H}$. In the proposed approach, through the generation of strata-wise samples, $\boldsymbol{\Phi} | \mathbb{S}_i$, and the corresponding site-specific wind speed $\hat{v}_H$, strata-wise construction of $f_{\hat{v}_H | \mathbb{S}_i}$ (or equivalently, the conditional cumulative distribution function) is enabled. Subsequently, these empirical quantities are combined with $\tilde{P}(\mathbb{S}_i)$, which is also estimated in the process, to obtain the hazard curve. In this example, the following holds $\pmb{\upsigma} = \boldsymbol{\Phi}$. 

While the evaluation of the hazard model, $\mathcal{H}$, is less computationally intensive than the nonlinear dynamic analysis involved in the response estimation, its computational cost is large enough to preclude the direct use of MC to generate strata-wise samples. The non-straight and non-stationary Gaussian stochastic wind load model outlined in \cite{ouyang2021performance} was adopted and calibrated to building-specific wind tunnel data to convert wind speed and direction time histories to stochastic aerodynamic floor loads through spectral proper orthogonal decomposition \citep{ChenPOD2005}. The time-varying wind loads complying with the hurricane evolution in the sub-region span several hours in duration. 

\subsubsection{Building system}

\begin{figure}
\centering
\includegraphics[scale = 0.64]{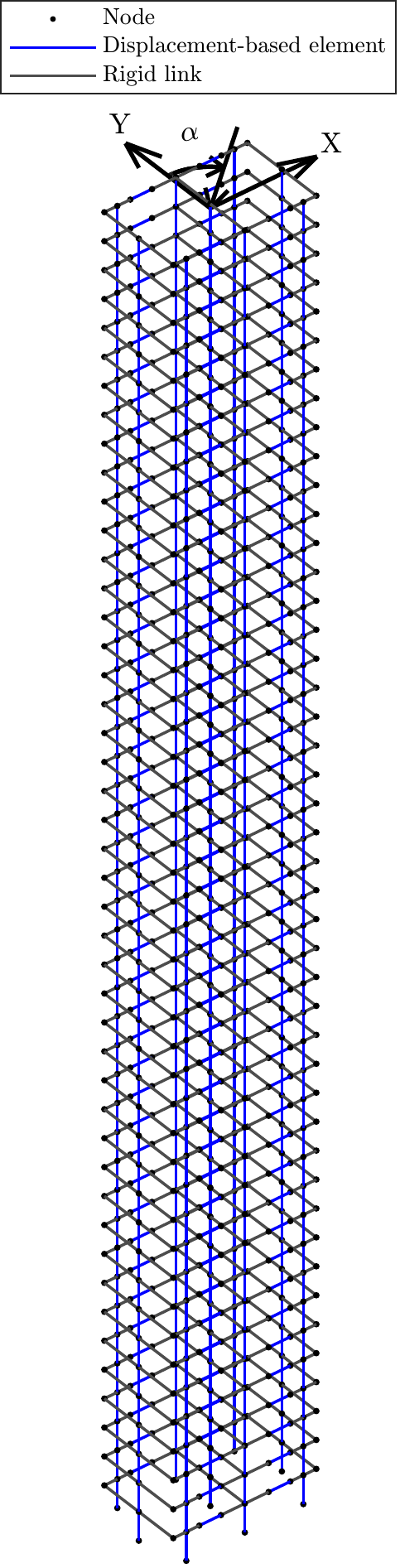}
\caption{Three-dimensional numerical model of the 45-story concrete building using equivalent frame method.}
\label{fig:45RCbuilding}
\end{figure}

The 45-story RC core building was designed by the ASCE 7-22 task committee on performance-based wind engineering. The lateral load resisting system is composed of multiple shear walls connected by coupling beams at each floor level. The shear walls were modeled using the equivalent frame method as columns modeled with displacement-based beam-column elements and rigid links whereas the floors were modeled as rigid diaphragms for horizontal movements. Figure \ref{fig:45RCbuilding} shows the structural model of the building. A modal damping ratio of 2\% was considered. A stress-resultant plasticity model was developed and solved through an adaptive fast nonlinear analysis (AFNA) scheme \citep{li2022rapid,li2021adaptive}. The approach captures second-order P-Delta effects through a linearized P-Delta model. Three-dimensional piece-wise linear yield surfaces were adopted for representing the yield domains of the reinforced concrete members, the details of which can be found in \cite{li2022rapid}. No system uncertainties were considered and the mean values reported in \cite{li2022rapid} were adopted for the material properties and gravity loads. It should be noted that not considering system uncertainties was simply a modeling choice in this case study and should not be viewed as a limitation of the proposed scheme. Here, $\pmb{\uptau}$ consists of the high-dimensional stochastic sequence (in the order of tens of thousands of random variables) within the stochastic wind load model enabling the capture of record-record variability.

\begin{figure}
\centering
\includegraphics[width = \textwidth]{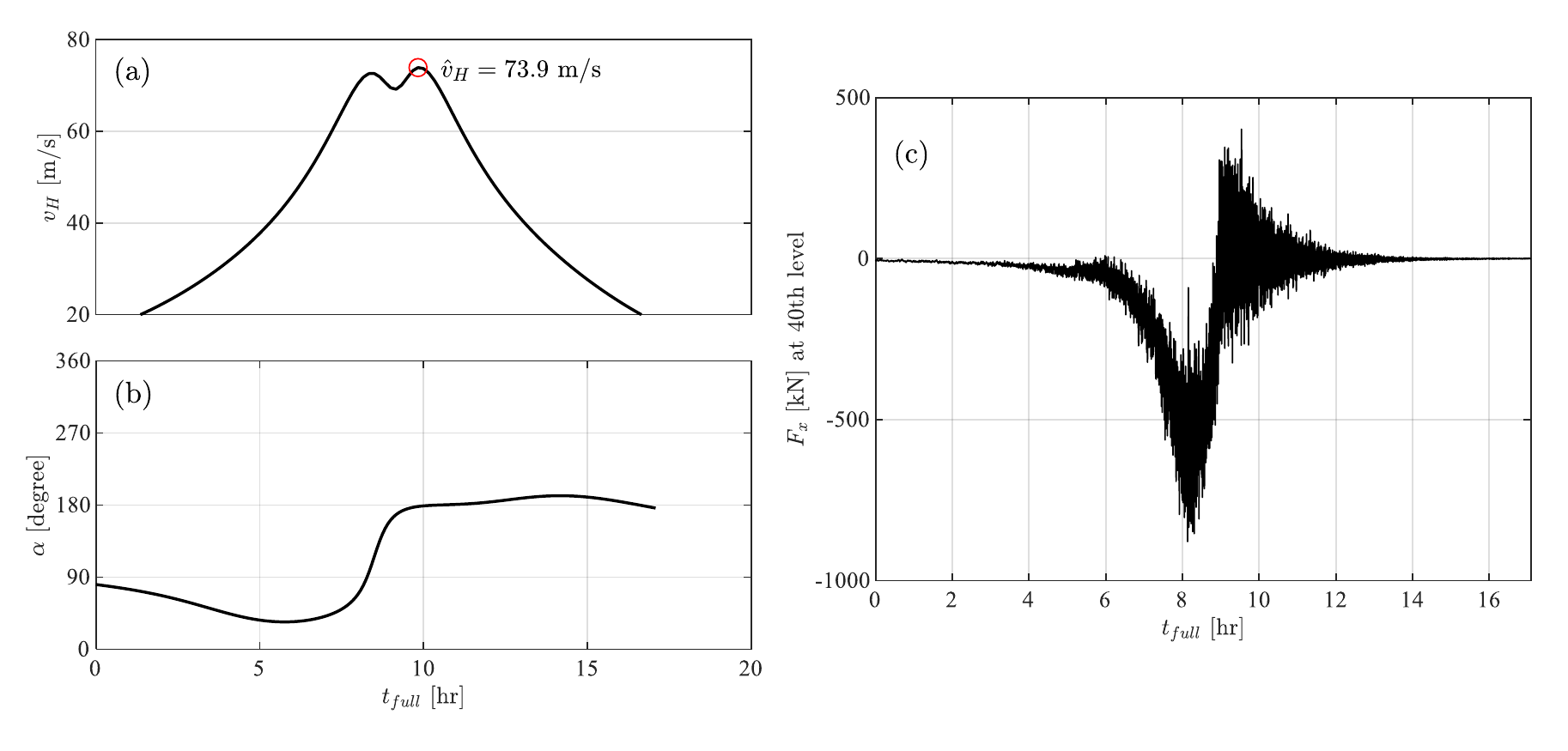}
\caption{Wind loading in a representative sample in the last stratum: (a) Evolution of the hourly mean wind speed; (b) wind direction; (c) X-direction wind load at the 40th level.}
\label{fig:Wload}
\end{figure}
\begin{figure}
\centering
\includegraphics[scale = 0.7]{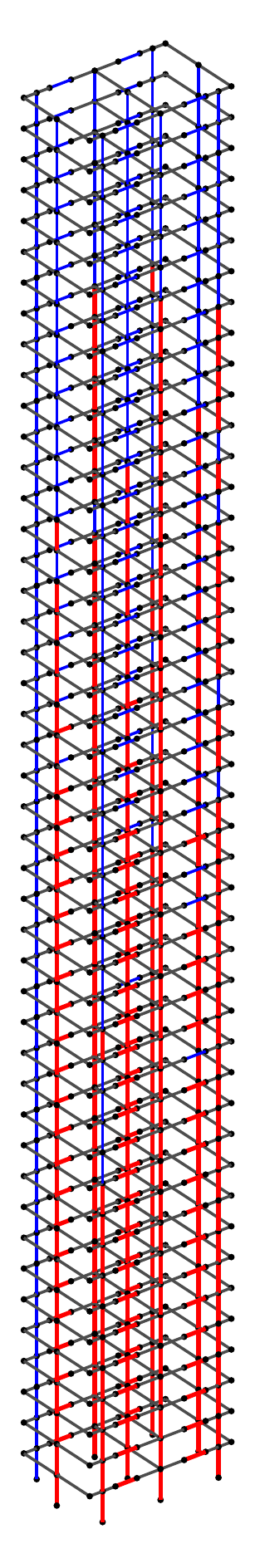}
\caption{Wind-induced structural yielding in a representative sample in the last stratum.}
\label{fig:YieldStructure}
\end{figure}

\subsubsection{Results}

\begin{figure}
\centering
\includegraphics[scale=0.85]{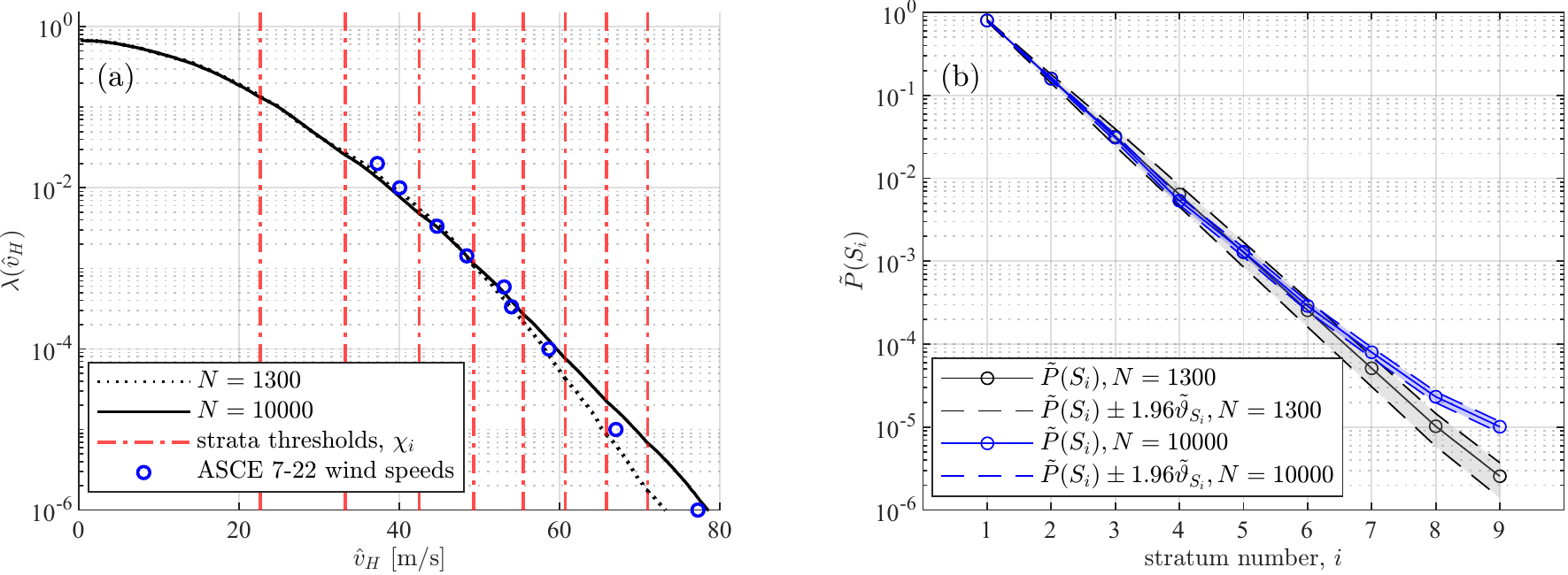}
\caption{(a) Updated wind speed hazard curve; (b) updated strata probabilities.}
\label{fig:hazardcurve}
\end{figure}

\begin{figure}
\centering
\includegraphics[width = 0.6\textwidth]{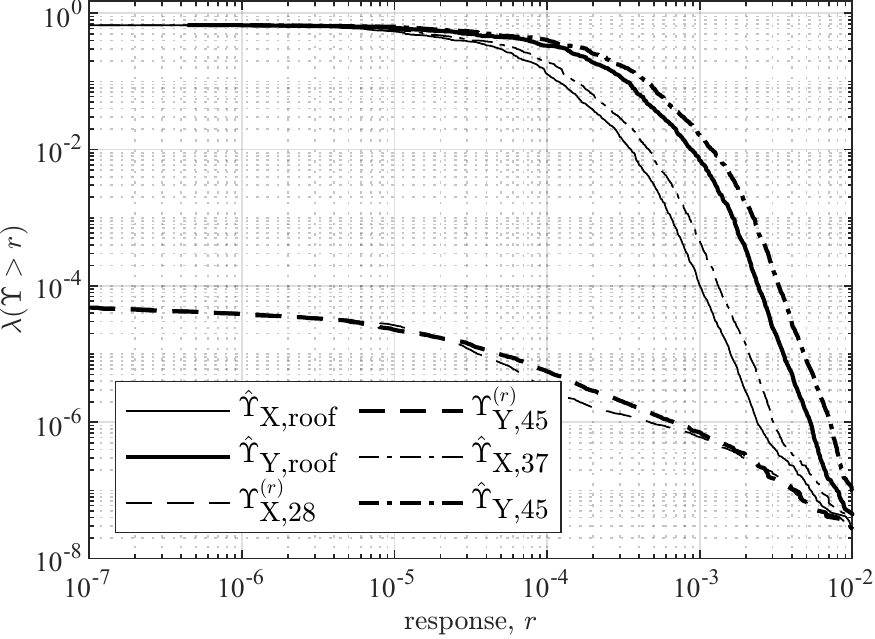}
\caption{Annual exceedance rate curves for the limit state functions.}
\label{fig:aercurves}
\end{figure}

For estimating the small failure probabilities, the construction of strata with low probabilities is essential, and therefore, to initialize the process with SuS-based Phase-I sampling, $m=9, p=0.2$ and $N=1300$ were considered. This ensured 1300 samples in the last stratum with $\tilde{P}(\mathbb{S}_m) = 0.2^8 = 2.56 \times 10^{-6}$. The SuS-based procedure took about four minutes to generate 9620 samples when run sequentially on an Intel i7-7700 3.60 GHz processor and for comparison, MC-based Phase-I sampling would have taken more than a month given $\mathscr{C}(\mathcal{H}) \approx 6$ milliseconds. For the preliminary study, $n_p = 150$ was considered, and the c.o.v targets, $\omega_h$ were set to 10\% only for the limit states associated with $\hat{\Upsilon}_{\text{Y,roof}}$ and $\hat{\Upsilon}_{\text{Y}}$. The largest peak IDRs were most often observed at the 37th story in the X direction and at the 45th story in the Y direction. Similarly, the largest residual IDRs were most often observed at the 28th story in the X direction and at the 45th story in the Y direction. For a representative sample in the last stratum, the time-varying wind speed, and direction are shown in Figure \ref{fig:Wload} corresponding to a 17-hour storm. The peak hourly-mean wind speed is also indicated in Figure \ref{fig:Wload}(a) and the resulting X-direction load at the 40th level is shown in Figure \ref{fig:Wload}(c). Corresponding to this large intensity of loading, the structure experiences significant nonlinearity that is illustrated by Figure \ref{fig:YieldStructure} where the considerable proportion (of about 56\%) of yielded elements (in red) at the end of the wind event is noteworthy. The generation of $mn_p = 1350$ response samples involved nonlinear dynamic analyses taking around 10 days to compute. Based on the preliminary study results, it was observed that due to the significant sample-allocation-independent variance contribution, the c.o.vs could not be reduced to less than about 20\%. This implies that $N = 1300$ constructs the hazard curve and estimates strata probabilities with large uncertainty that is inadequate for attaining the target c.o.vs. Therefore, Phase-I sampling was repeated with $N = 10,000$ but with fixed strata thresholds as given by the previous trial. It was expected that the lower limit of the c.o.vs would approximately reduce by a factor of $\sqrt{\frac{1300}{10000}}$ and could be brought down to less than 10\%. Notably, the time taken to repeat the Phase-I sampling was only about 20 minutes. Figure \ref{fig:hazardcurve}(a) compares the hazard curves constructed using $N=1300$ and $N=10,000$, as well as indicates the division of the wind speed range that reflects the stratification. The difference is significant in the large wind speed range as a result of successively accumulating errors in the case of $N = 1300$. The site-specific ASCE 7-22 wind speeds \citep{ASCE7} are also reported. Figure \ref{fig:hazardcurve}(a) also illustrates how through the application of the SuS-based Phase-I sampling, large wind speeds at the tail of its distribution could be efficiently sampled to enable a direct simulation of extreme structural responses. The correspondence between Figure \ref{fig:hazardcurve}(a) and Figure \ref{fig:sampledistribution_illus}(b) is also worth mentioning. More importantly, Figure \ref{fig:hazardcurve}(b) shows the update in the strata probabilities, including the estimation error, wherein the shaded region indicates a scatter of 1.96 times the standard deviation, $\tilde{\vartheta}_{\mathbb{S}_i}$, around the estimates. 
The updated strata probabilities and the results of the optimization are reported in Table \ref{tab:optimalsample}. Although the estimate $\tilde{P}(\mathbb{S}_m)$ has increased, the c.o.v in its estimation dropped from 23.0\% to 7.9\%, roughly by the factor $\sqrt{\frac{1300}{10000}}$. For the three limit states considered in the optimization procedure, additional samples, $(n_i - n_p)$, were required only in the last three strata.
\begin{table}
\caption{Stratification and optimal sample allocation.}
\footnotesize
\centering
\begin{tabular}{ccccc} \\\hline 
Stratum & $\chi_{L}$ [m/s] & $\chi_{U}$ [m/s] & $\tilde{P}(\mathbb{S}_i)$ & $n_i$ \\\hline
Stratum 1 & 0.00 & 22.63 & $8.04\times 10^{-1}$ & 150\\
Stratum 2 & 22.63 & 33.21 & $1.58\times 10^{-1}$ & 150\\
Stratum 3 & 33.21 & 42.45 & $3.10\times 10^{-2}$ & 150\\
Stratum 4 & 42.45 & 49.23 & $5.38\times 10^{-3}$ & 150\\
Stratum 5 & 49.23 & 55.43 & $1.30\times 10^{-3}$ & 150\\
Stratum 6 & 55.43 & 60.66 & $2.90\times 10^{-4}$ & 150\\
Stratum 7 & 60.66 & 65.82 & $7.99\times 10^{-5}$ & 170\\
Stratum 8 & 65.82 & 70.96 & $2.33\times 10^{-5}$ & 514\\
Stratum 9 & 70.96 & $\infty$ & $1.01\times 10^{-5}$ & 1146
\\\hline
\end{tabular}
\label{tab:optimalsample}     
\end{table}
The annual failure probabilities, $\tilde{P}_{f,h}$, for all eight limit states were estimated using a total of $n = 2730$ response evaluations. Since these probabilities are conditional on the occurrence of a hurricane event, they were transformed into annual exceedance rates (AERs) by multiplying with $\lambda_{\text{hurr}}$. The AERs, the associated c.o.vs and the 50-year reliability indices, estimated as $\beta_{50} = \Phi_N^{-1}[(1-\lambda_{\text{hurr}}\tilde{P}_{f,h})^{50}]$ where $\Phi_N$ is the standard normal distribution function, are reported in Table \ref{tab:finalresults}. Clearly, the c.o.vs for the limit states LS2, LS4, and LS8 are around 10\% as targeted and demonstrate the capability of the proposed procedure to achieve a desired level of confidence in the estimates. The enormous efficiency gain provided by the procedure can be better appreciated by observing that for attaining the c.o.vs reported in Table \ref{tab:finalresults}, a simple MC simulation would have required samples in the range of $n_{\text{MC}} \approx 10^4 n$ for all limit states except LS8, which would have required $\approx 10^3 n$ samples and LS4 which would have required $\approx 10^2 n$ samples. In other words, and as illustrated in Table \ref{tab:finalresults} through the ratio $n_{\text{MC}}/n$, a reduction of several orders of magnitude in necessary samples for achieving a target accuracy is achieved through the application of the proposed approach. The AER curves for the quantities of interest as a function of the response values can also be constructed, similar to the hazard curve, through the total probability theorem and are reported in Figure \ref{fig:aercurves}. This figure also highlights how the proposed scheme enabled the simulation of extreme responses associated with small annual exceedance rates. It can be noted from these curves that, in general, the Y-direction responses are more dominant for the structure relative to X.

\begin{table}
\caption{Annual failure rates and estimation error for example 1.}
\footnotesize
\centering
\begin{tabular}{cccccc} \\\hline
Limit states & Description & AER & $\beta_{50}$ & c.o.v & $n_{\text{MC}}/n$ \\\hline
LS1 & $\hat{\Upsilon}_{\text{X,roof}} > 1/200$  & $1.24 \times 10^{-7}$ & 4.37 & 24.0\% & $3.45 \times 10^4$\\
LS2 & $\hat{\Upsilon}_{\text{Y,roof}} > 1/200$ & $1.43 \times 10^{-6}$ & 3.80 & 11.2\% & $1.37 \times 10^4$\\
LS3 & $\hat{\Upsilon}_{\text{X,roof}} > 1/400$ & $8.52 \times 10^{-7}$ & 3.93 & 12.8\% & $1.75 \times 10^4$\\
LS4 & $\hat{\Upsilon}_{\text{Y,roof}} > 1/400$ & $8.04 \times 10^{-5}$ & 2.65 & 9.6\% & $3.32 \times 10^2$\\
LS5 & $\Upsilon^{(r)}_{\text{X,28}} > 1/1000$ & $6.15 \times 10^{-7}$ & 4.01 & 14.7\% & $1.84 \times 10^4$\\
LS6 & $\Upsilon^{(r)}_{\text{Y,45}} > 1/1000$ & $7.09 \times 10^{-7}$ & 3.97 & 12.4\% & $2.25 \times 10^4$\\
LS7 & $\hat{\Upsilon}_{\text{X,37}} > 1/200$ & $2.19 \times 10^{-7}$ & 4.24 & 18.3\% & $3.37 \times 10^4$\\
LS8 & $\hat{\Upsilon}_{\text{Y,45}} > 1/200$ & $8.21 \times 10^{-6}$ & 3.35 & 10.7\% & $2.63 \times 10^3$
\\\hline
\end{tabular}
\label{tab:finalresults}     
\end{table}

\subsection{Example 2: Ground Motion-excited Steel Frame}
\subsubsection{Overview}
In this example, the objective is to estimate multiple failure probabilities associated with IDR-based limit states for a four-story archetype structure subjected to stochastic ground motions. The spectral acceleration at the first-mode period with 5\% damping, $S_a(T_1, 5\%)$, is selected as the stratification variable. Unlike the peak ground acceleration (PGA), which is only a characteristic of the ground motion, spectral acceleration also accounts for the frequency content of the excitation around the structure’s first-mode period \citep{jalayer2006information}. It is a popularly used intensity measure (IM) in seismic risk analysis. The choice of $\chi = S_a$ is motivated by the expectation that the variability in nonlinear responses at a given value of $S_a$ is much less than that in the entire response set \citep{shome1998earthquakes}. The following 12 limit states are considered: structural collapse, defined as maximum peak IDR exceeding 15\% \citep{elkadygithub}; peak IDR for each of the four stories, $\hat{\Upsilon}_k, 1\leq k\leq 4$, and its maximum (over all stories) exceeding 3\%; residual IDR for each story, $\Upsilon_k^{(r)}, 1\leq k\leq 4$, and its maximum (over all stories) exceeding 1.41\%; and finally, residual roof drift ratio, $\Upsilon^{(r)}_{\text{roof}}$, exceeding 0.91\%. The thresholds for the peak, residual IDRs, and residual roof drift ratio are selected on the basis of repairability limits suggested in literature \citep{iwata2006reparability,bojorquez2013residual}.

\subsubsection{Stochastic ground motion model}

A point-source stochastic model is adopted for ground motion modeling where the spectrum of the ground motion that encapsulates both the physics of the fault rupture, as well as the wave propagation, is expressed as a product of the source, $E(f;M)$, path, $P(f; r)$, and site, $G(f)$, contributions \citep{boore2003simulation}. The frequency-dependent total spectrum, $A(f; M,r)$, is parameterized by the seismic moment magnitude, $M$, and epicentral distance, $r$, to characterize the seismic hazard. That is,
\begin{equation}
    A(f; M,r) = (2 \pi f)^2 E(f; M) P(f; r) G(f)
\end{equation}
In particular, the two-corner point-source model developed by \cite{atkinson2000stochastic} for California sites is used, wherein the functional form of the source spectrum contains two corner frequencies. The duration of the ground motion is determined by the time-dependent envelope function, $e(t; M,r)$, which is yet again parameterized by $M$ and $r$. Ultimately, the ground motion acceleration time history is generated according to this model by modulating a white noise sequence, $\textbf{Z}$, by $e(t; M,r)$, transforming into the frequency domain, normalizing it before multiplying by $A(f; M,r)$ and finally transforming it back to the time domain \citep{boore2003simulation}. The high-dimensional vector $\textbf{Z}$ models the record-record variability while the uncertain seismic hazard parameters, $M$ and $r$, represent the dominant risk factors \citep{vetter2012global}. The predictive relationships that relate the source, path, and site contributions, as well as the time-domain envelope function to $M$ and $r$, can be found elsewhere \citep{atkinson2000stochastic,boore2003simulation,vetter2012global}. In calibrating the ground motion model, the following parameters were adopted: Radiation pattern $R_\Phi = 0.55$, source shear-wave velocity $\beta_s = 3.5$ km/s, density $\rho_s = 2.8$ g/cm$^3$, seismic velocity $c_Q = 3.5$ km/s; an elastic attenuation factor $Q(f) = 180f^{0.45}$ (for California region according to \cite{atkinson2000stochastic}), geometric spreading function $Z(R) = 1/R$ for $R<70$ km and $Z(R) = 1/70$ for $R>=70$ km, where $R$ is the radial distance from the source to site; the path-independent energy loss is modeled by the diminution function which is expressed by the $f_{max}$ filter, where $f_{max} = 15$ rad/s; finally, the site amplification is described for NEHRP ``D'' site condition (i.e., the building site condition) using empirical curves presented in \cite{boorejoyner1997site}. The duration of the simulated stochastic ground accelerations is 60 s with $\Delta t = 0.01$ s. Therefore, the length of $\textbf{Z}$ is 6001. The parameters $\lambda_t$ and $\eta_t$ in the envelope function were set to 0.2 and 0.05, respectively, as suggested in \cite{boore2003simulation}.

The moment magnitude $M$ was modeled by the bounded Gutenberg-Richter recurrence relationship as a truncated exponential distribution with $M_{min} = 6$ and $M_{max} = 8$ \citep{kramer_2003}:
\begin{equation}
    \label{eq:GRrelation}
    p(M) = \frac{\beta \text{ exp}(-\beta (M-M_{min}))}{1-\text{exp}(-\beta (M_{max} - M_{min}))} \quad M_{min} \leq M \leq M_{max} 
\end{equation}
where the regional seismicity factor $\beta$ is chosen as $0.9\log_e(10)$. Eq. \eqref{eq:GRrelation} could equivalently be expressed as an equation for the mean annual rate of exceedance, $\lambda_M$, of an earthquake of magnitude $M$ by setting a value for the exceedance rate for the lower threshold magnitude, $\lambda_{M_{min}}$ \citep{kramer_2003}. In this study, $\lambda_{M6} = 0.6$. The uncertainty in $r$ is modeled using a lognormal distribution with median of 15 km and c.o.v of 0.4. Here, $\pmb{\upsigma} = [M,r,\textbf{Z}]$ and the function $\mathcal{H}$ involved in computing $S_a$ is the ground motion model evaluation followed by the linear oscillator response estimation which combined only takes 3-4 milliseconds to run sequentially on an Intel i7-7700 3.60 GHz processor.

\subsubsection{Building description}

\begin{figure}[b]
\centering
\includegraphics[width = 0.6\textwidth]{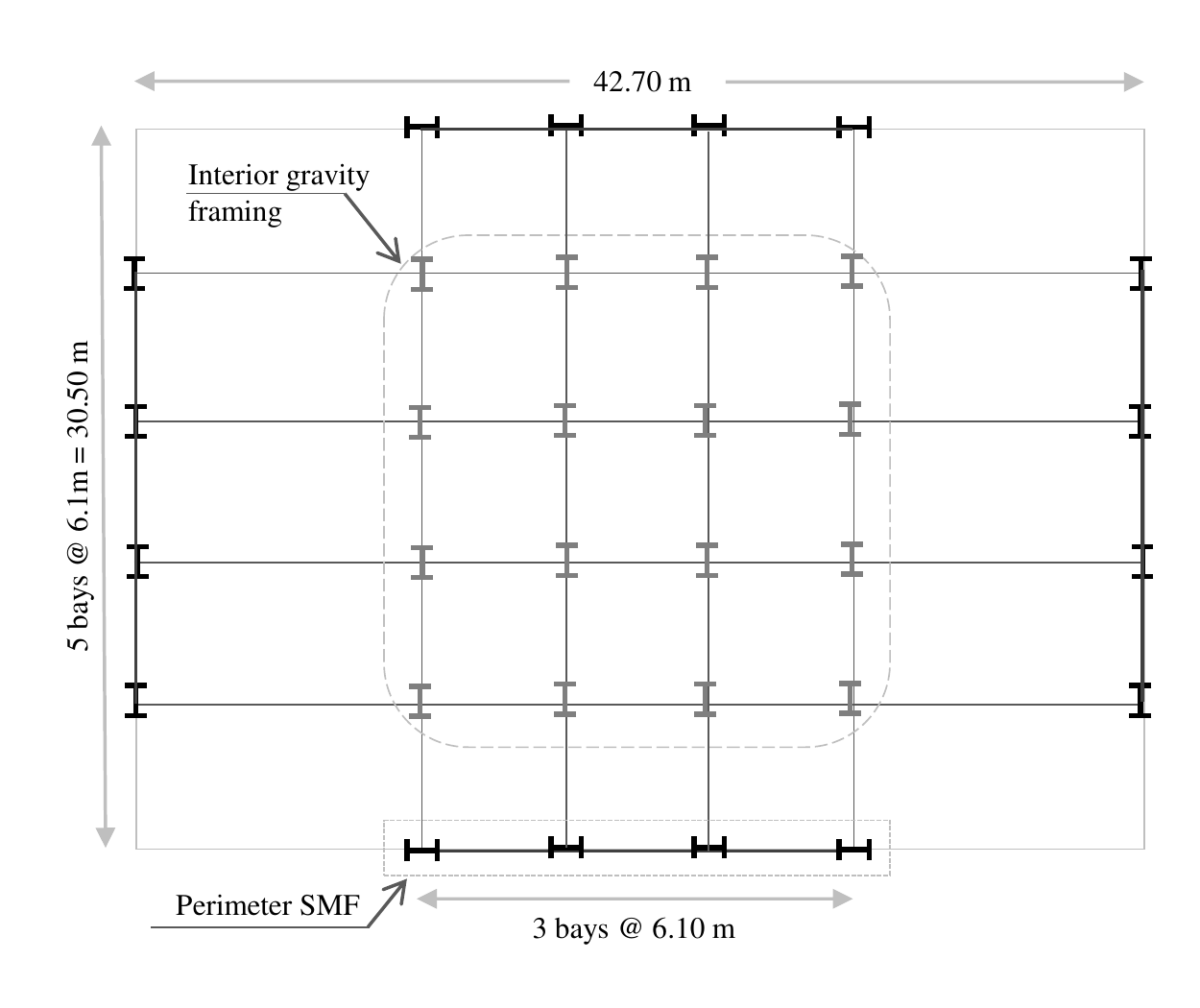}
\caption{Plan view of the four-story archetype steel building.}
\label{fig:4storybuilding}
\end{figure}

A four-story archetype office steel building designed with perimeter special moment frames (SMFs) assumed to be located in downtown Los Angeles, California, is considered in this study. The schematic plan view of the building is shown in Figure \ref{fig:4storybuilding}. 
The considered two-dimensional nonlinear model (noted as the ``B model'' in \cite{elkady2015effect}) represents the building in the E-W loading direction. It models the bare steel structural components of the SMF while ignoring the effects of both the composite floor slab and the gravity framing. This model was developed by \cite{Elkadythesis,elkadygithub} in the Open System for Earthquake Engineering Simulation (OpenSees) platform \citep{OpenSees}. The fundamental period of the structure, $T_1$, is $1.43$ s and the building height $H$ is $16.5$ m. The key modeling aspects include panel zone modeling, reduced-beam-section connections, consideration of P-Delta effects using a fictitious ``leaning'' column, and member modeling using a combination of elastic elements and flexural springs at their ends. Rayleigh damping is calibrated by assigning the damping ratios, $\zeta$, of the first and third modes. The material yield strength, $F_y$, and $\zeta$ are modeled as lognormal random variables with a mean of 417 MPa and 1.5\%, respectively, and c.o.v of 0.06 and 0.4. Here, $\pmb{\uptau} = [F_y,\zeta]$.


\subsubsection{Results}

\begin{figure}[b]
\centering
\includegraphics[width = 0.55\textwidth]{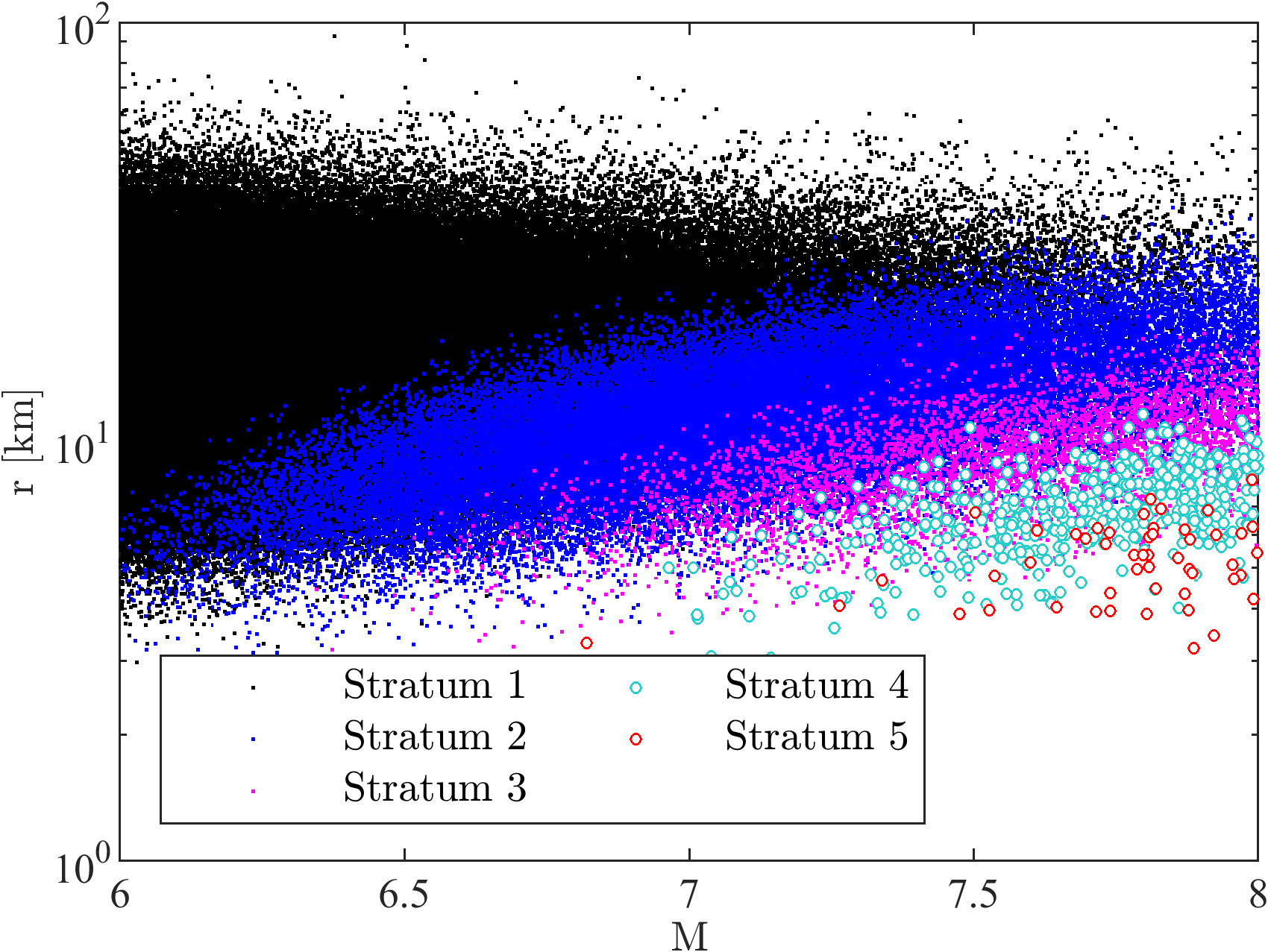}
\caption{Strata-wise sample scatter of $M$ and $r$.}
\label{fig:Mrdistribution}
\end{figure}

\begin{figure}
\centering
\includegraphics[width = 0.55\textwidth]{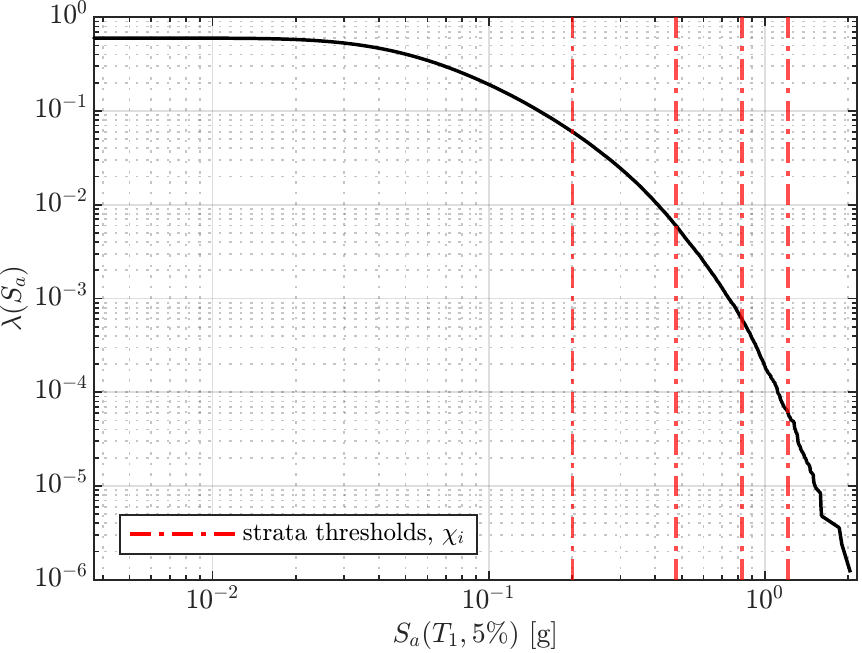}
\caption{Spectral acceleration hazard curve.}
\label{fig:speccurve}
\end{figure}

\begin{figure*}[]
\centering
\includegraphics[scale=0.85]{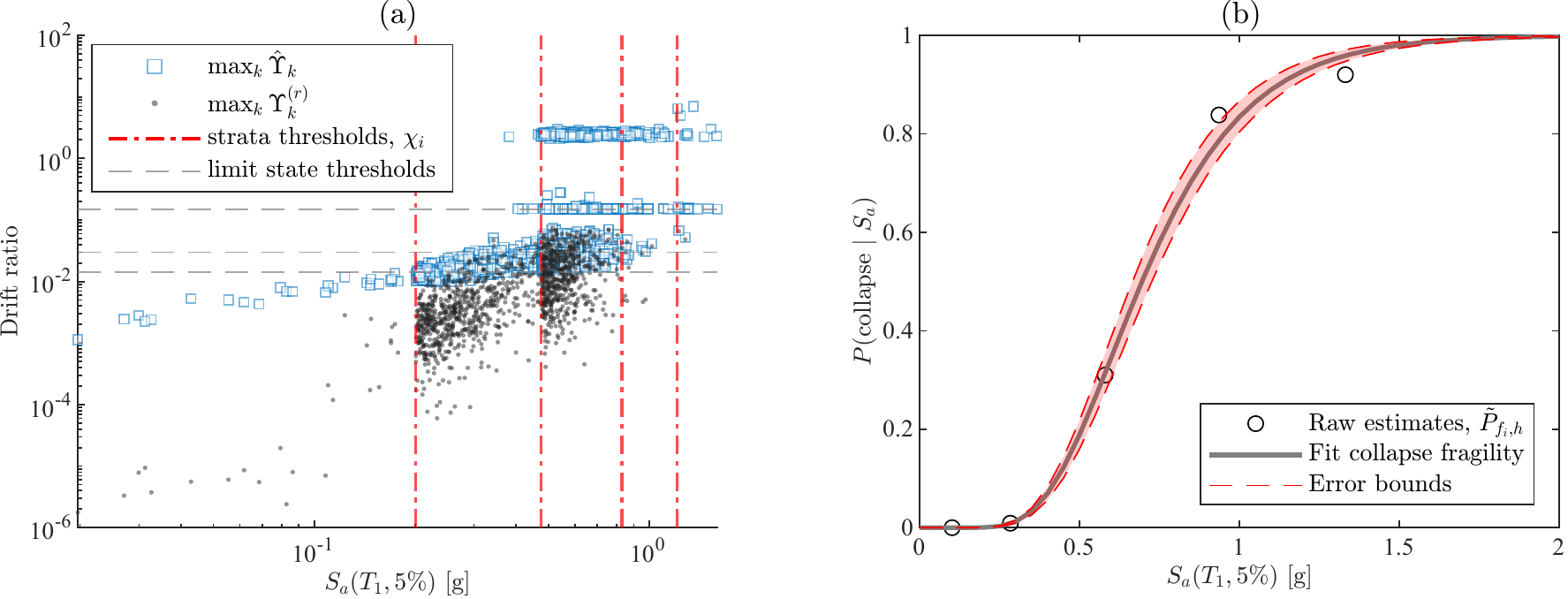}
\caption{(a) Evolution of drift ratios with increasing $S_a (T_1,5\%)$ and stratum number; (b) estimated collapse fragility curve with error bounds.}
\label{fig:collapsefragility}
\end{figure*}

\begin{figure}
\centering
\includegraphics[width = 0.65\textwidth]{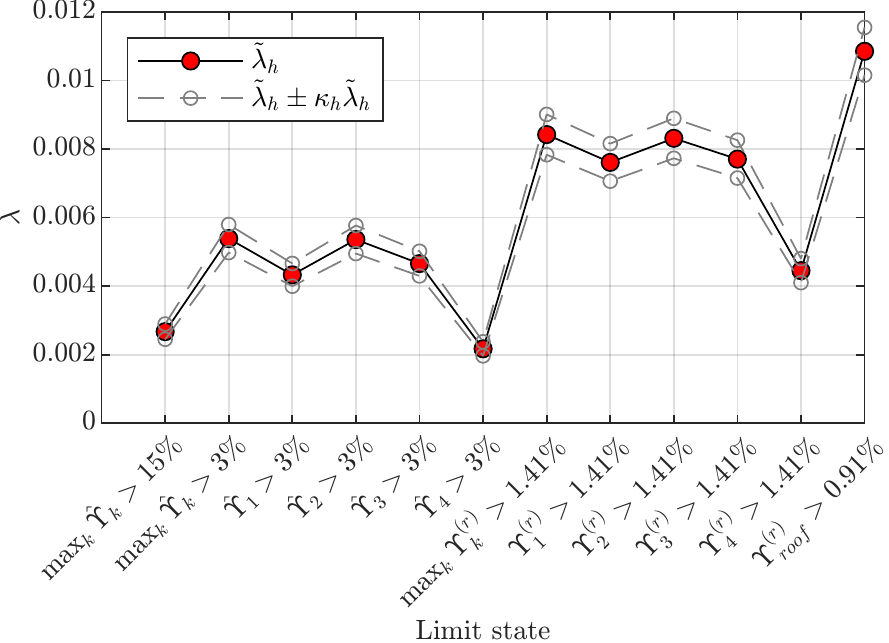}
\caption{AER with error estimation.}
\label{fig:GMAER}
\end{figure}

Since $\mathscr{C}(\mathcal{H})$ is negligible, MC-based Phase-I sampling is considered. By setting, $m=5$, $p=0.1$, and $\hat{n}=5 \times 10^5$, it took around 30 minutes to generate enough samples to have 50 in the last stratum with $\tilde{P}(\mathbb{S}_5) = 10^{-4}$. The strata boundaries were adaptively obtained as $\{\chi_0,\chi_1,\chi_2,\chi_3,\chi_4,\chi_5\} = \{0,0.20,0.48,0.83,1.21,\infty\}$ in units of $g$ (acceleration due to gravity). Figure \ref{fig:Mrdistribution} shows the strata-wise sample scatter of the seismic hazard parameters, $M$ and $r$. 
The figure illustrates the well-known downside of MC sampling which is the wasteful generation of abundant Phase-I samples in the earlier strata, roughly in proportion to the strata probabilities, in order to generate the required number in the last stratum. For the preliminary study, $n_p = 25$ was considered and the c.o.v targets, $\omega_h$ were set to 10\% for all limit states, except for collapse for which it was set to 5\%. The results of the preliminary study for certain key limit states and the estimated optimal sample sizes are reported in Table \ref{tab:GMoptimalsample}. 
\begin{table*}[t]
\caption{Preliminary study and optimal sample allocation.}
\tabcolsep 4pt                                                          
\footnotesize
\centering
\begin{tabular}{ccccccc} \\\hline
Stratum & $\tilde{P}(\mathbb{S}_i)$ & $\tilde{P}(\max_k \hat{\Upsilon}_k > 15\%)$ & $\tilde{P}(\max_k \hat{\Upsilon}_k > 3\%)$ & $\tilde{P}(\max_k \Upsilon_k^{(r)} > 1.41\%)$ & $\tilde{P}(\Upsilon^{(r)}_{\text{roof}} > 0.91\%)$ & $n_i$ \\\hline
Stratum 1 & $9\times10^{-1}$ & 0 & 0 & 0 & 0 & 25\\
Stratum 2 & $9\times10^{-2}$ & 0 & 0 & 0.04 & 0.12 & 659\\
Stratum 3 & $9\times10^{-3}$ & 0.24 & 0.52 & 0.52 & 0.60 & 797\\
Stratum 4 & $9\times10^{-4}$ & 0.84 & 0.92 & 0.96 & 0.96 & 68\\
Stratum 5 & $1\times10^{-4}$ & 0.92 & 1.00 & 1.00 & 1.00 & 25
\\\hline
\end{tabular}
\label{tab:GMoptimalsample}     
\end{table*}
Notably, no additional samples were needed in $\mathbb{S}_5$ since the refined estimation of conditional probabilities in the earlier strata with higher strata probabilities was preferred by the optimization algorithm to meet the c.o.v targets. It was found that the total number of simulations required is $n = 1574$ inclusive of the 125 preliminary test samples. Figure \ref{fig:speccurve} shows the estimated spectral acceleration hazard curve along with the strata thresholds. 
Figure \ref{fig:collapsefragility}(a) illustrates the evolution of $\max_k \hat{\Upsilon}_k$ and $\max_k \Upsilon_k^{(r)}$ with different intensity levels of $S_a (T_1,5\%)$ and the stratum number. The increasing trend of the drift ratios with increasing spectral acceleration provides support for the choice of the latter as the stratification variable. It should be noted that the figure only shows the maximum residual IDRs for the non-collapse samples as they cannot be quantified for the collapse samples, however, the corresponding limit states are assumed to be violated. The procedure also enables a natural construction of fragility functions when the pertinent hazard intensity measure is selected as the stratification variable. For instance, lognormal collapse fragility can be defined by first assuming each point estimate of the conditional collapse probabilities to be located at the average $S_a (T_1,5\%)$ in the associated stratum, and secondly, by applying the maximum likelihood approach for fitting \citep{baker2015}. Additionally, the calculation of $\tilde{\vartheta}_{i,h}$ for collapse enables the specification of error bounds for the fragility curves. Following this approach, Figure \ref{fig:collapsefragility}(b) reports the collapse fragility curve and error confidence bounds estimated with the conditional probabilities set to $\tilde{P}_{f_i,h} \pm 1.65\tilde{\vartheta}_{i,h}$. The median of the collapse fragility curve is 0.69 g, while the dispersion is 0.37. Finally, the overall failure probabilities were estimated and multiplied by $\lambda_{M6} = 0.6$ to convert to AERs. The AERs and their c.o.vs are expressed both graphically in Figure \ref{fig:GMAER} as well as in Table \ref{tab:GMfinalresults}. Figure \ref{fig:GMAER} reports the estimated AERs along with the error margins represented using their standard deviation, $\kappa_h \tilde{\lambda}_h$. Clearly, the target accuracy in the estimations have been met for all limit states except for collapse. The violation of this c.o.v target can be attributed to the fact that the 5\% target was an active constraint in the optimization procedure, therefore, more sensitive to the accuracy of the preliminary-study-based optimal sample sizes. However, the preliminary study incorrectly estimated the conditional collapse probability for $\mathbb{S}_2$ of about 0.91\% to be zero which also highlights the importance of $n_p$. It would be reasonable to specify stricter c.o.v targets than desired at the stage of optimization if they are to be met rigorously, although it may potentially increase the sampling demands. Notably, the relatively large annual failure rates in this case study justify the use of $m=5$ and only $n=1574$ samples for providing estimations with high accuracy. As would be expected, the variance reduction factor, $n_{\text{MC}}/n$, is more modest than seen in the first example, although still in the order of one magnitude, due to the relatively large failure rates in comparison to those of the first example.

\begin{table}
\caption{Annual failure rates and estimation error for example 2.}
\footnotesize
\centering
\begin{tabular}{ccccc} \\\hline
Limit states & Description & AER & c.o.v & $n_{\text{MC}}/n$ \\\hline
LS1 & $\max_k \hat{\Upsilon}_k > 15\%$  & $2.67 \times 10^{-3}$ & 8.32\% & 20.5\\
LS2 & $\max_k \hat{\Upsilon}_k > 3\%$ & $5.39 \times 10^{-3}$ & 7.59\% & 12.2\\
LS3 & $\hat{\Upsilon}_1 > 3\%$ & $4.33 \times 10^{-3}$ & 7.66\% & 14.9\\
LS4 & $\hat{\Upsilon}_2 > 3\%$ & $5.36 \times 10^{-3}$ & 7.63\% & 12.1\\
LS5 & $\hat{\Upsilon}_3 > 3\%$ & $4.66 \times 10^{-3}$ & 7.72\% & 13.6\\
LS6 & $\hat{\Upsilon}_4 > 3\%$ & $2.17 \times 10^{-3}$ & 9.43\% & 19.7\\
LS7 & $\max_k \Upsilon_k^{(r)} > 1.41\%$ & $8.42 \times 10^{-3}$ & 6.98\% & 9.2\\
LS8 & $\Upsilon_1^{(r)} > 1.41\%$ & $7.61 \times 10^{-3}$ & 7.19\% & 9.6\\
LS9 & $\Upsilon_2^{(r)} > 1.41\%$ & $8.31 \times 10^{-3}$ & 7.01\% & 9.2\\
LS10 & $\Upsilon_3^{(r)} > 1.41\%$ & $7.70 \times 10^{-3}$ & 7.17\% & 9.5\\
LS11 & $\Upsilon_4^{(r)} > 1.41\%$ & $4.45 \times 10^{-3}$ & 7.87\% & 13.7\\
LS12 & $\Upsilon^{(r)}_{\text{roof}} > 0.91\%$ & $1.08 \times 10^{-2}$ & 6.42\% & 8.4
\\\hline
\end{tabular}
\label{tab:GMfinalresults}     
\end{table}

\section{Conclusion}
The evaluation of extreme nonlinear structural responses using complex models and the description of the uncertainty in the exceedance of associated acceptance criteria using failure probabilities has become central to modern performance-based engineering approaches. Building on the idea of classic double sampling, in this paper an extended two-phase-sampling-based stochastic simulation scheme is proposed to tackle high-dimensional reliability problems in natural hazard applications characterized by multiple limit states. The proposed methodology is cast as a generalization of stratified sampling wherein the Phase-I sampling generates strata-wise samples and estimates the strata probabilities. Phase-I sampling enables the selection of a generalized stratification variable for which the probability distribution is not known \textit{a priori}. To improve the efficiency, Phase-I sampling takes the form of SuS when the use of MC is deemed infeasible. Notably, the first case study illustrated the significance of Phase-I sampling in realizing the adequate accuracy in the estimated strata probabilities, which in turn affected the attainable lower limit on the c.o.vs. The benefits of employing SuS over MC are tremendous when the Phase-I sampling demands are high. On the other hand, the goal of Phase-II sampling is to estimate the final failure probabilities within the constraints of target c.o.vs with a minimum number of evaluations of the performance functions. This is achieved by an optimization approach that requires a preliminary simulation-based study as well as mathematical expressions for the c.o.vs. Therefore, the required expressions were derived while taking into account the sample correlations induced by MCMC and the uncertainty in the strata probabilities. The case study examples demonstrated not only the estimation of large reliabilities for multiple limit states with error measures but also the capability to roughly control these estimation errors with a minimum computational expense.

\section*{Acknowledgments}
This research effort was supported in part by the National Science Foundation (NSF) under Grant No. CMMI-1750339 and CMMI-2118488. This support is gratefully acknowledged.

\section*{Data Availability Statement}
All data, models, and code generated or used during the study appear in the submitted article.

\appendix
\section{Statistical properties of the SuS-based double sampling estimator}
\label{appdixA}

\subsection{Properties of $\tilde{P}_{f_i,h}$}
The modified version of the Metropolis-Hastings (M–H) sampler that was proposed by \cite{au2001originalsubset} is adopted in this study which is based on a component-wise sample generation to avoid the small acceptance rate of the original M–H sampler in high dimensions. Samples $\pmb{\uptheta}^{(i-1)}$ in $F_{i-1}, 2\leq i \leq m$ are distributed as $q(\pmb{\uptheta} | F_{i-1})$ and represent consecutive states of a Markov chain (typically, multiple chains exist arising from multiple seeds) with $q(\pmb{\uptheta} | F_{i-1})$ as the stationary distribution. A separate treatment of $\pmb{\uptau}$ is not necessary as it is independent of $\pmb{\upsigma}$ and therefore, unaffected by any conditioning on $\mathbb{S}_i$ (that is, $q(\pmb{\uptau} | \mathbb{S}_i) = q(\pmb{\uptau})$). Therefore, for simplicity of notation, $\pmb{\uptheta}$ is written with both $\pmb{\upsigma}$ (samples generated using subset simulation) and $\pmb{\uptau}$ (generated with MC simulation) included and not explicitly stated hereafter. It can be shown that the samples $\pmb{\uptheta}^{(i-1)}_k \in \mathbb{S}_i$ will be distributed as:
\begin{equation}
    \begin{split}
        \frac{\mathbbm{1}_{\mathbb{S}_i}(\pmb{\uptheta}| F_{i-1}) q(\pmb{\uptheta}| F_{i-1})}{P(\mathbb{S}_i | F_{i-1})} &= \frac{\mathbbm{1}_{\mathbb{S}_i}(\pmb{\uptheta}| F_{i-1}) q(\pmb{\uptheta}) \mathbbm{1}_{F_{i-1}}(\pmb{\uptheta})}{P(F_{i-1})P(\mathbb{S}_i | F_{i-1})} \\& =
        \frac{\mathbbm{1}_{\mathbb{S}_i}(\pmb{\uptheta}) q(\pmb{\uptheta})}{P(\mathbb{S}_i)} = q(\pmb{\uptheta}| \mathbb{S}_i)
    \end{split}
\end{equation}
This implies that $\mathbb{E}\left(\mathbbm{1}_{f,h}(\pmb{\uptheta}^{(i-1)}_k)\right) = P_{f_i,h}$  and consequently, $\mathbb{E}(\tilde{P}_{f_i,h}) = P_{f_i,h}$, where $\tilde{P}_{f_i,h}$ is the sample mean of the failure indicator function over a random subset of $n_i$ samples (denoted as $\mathcal{W}_i$) selected from the set of $\hat{n}_i$ samples (denoted as $\hat{\mathcal{W}}_i$) in $\mathbb{S}_i$ expressed as:
\begin{equation}
    \tilde{P}_{f_i,h} = \frac{1}{n_i} \sum_{k}\mathbbm{1}_{f,h}(\pmb{\uptheta}^{(i-1)}_k)
\end{equation}
The variance of $\tilde{P}_{f_i,h}$ can be derived using the following assumptions and notations similar to \cite{au2001originalsubset}: (a) At the $(i-1)$ simulation level, although the samples in $\mathcal{W}_i$ (and $\hat{\mathcal{W}}_i$) are in general dependent due to the seeds themselves being correlated, inter-chain correlation with respect to the occurrence of failure is assumed to be zero, i.e., $\mathbb{E}\left((\mathbbm{1}_{fh,jk}^{(i-1)} - P_{f_i,h})(\mathbbm{1}_{fh,j'k'}^{(i-1)} - P_{f_i,h})\right) = 0$ for  $j\neq j'$, where $\mathbbm{1}_{fh,jk}^{(i-1)}$ denotes $\mathbbm{1}_{f,h}(\pmb{\uptheta}^{(i-1)}_{jk})$ and $\pmb{\uptheta}^{(i-1)}_{jk} \in F_{i-1}$ denotes the $k$th sample in the $j$th Markov chain; and (b) the covariance between $\mathbbm{1}_{fh,jk}^{(i-1)}$ and $\mathbbm{1}_{fh,jk'}^{(i-1)}$ for the samples in $\mathbb{S}_i$ is denoted as:
\begin{equation}
\label{RSi_eqn}
    R_{\mathbb{S}_i} (k-k') = \mathbb{E}\left((\mathbbm{1}_{fh,jk}^{(i-1)} - P_{f_i,h})(\mathbbm{1}_{fh,jk'}^{(i-1)} - P_{f_i,h})\right)
\end{equation}
where the stationarity of the sample sequence is invoked and hence the dependency is only on the relative distance between the states ($k-k'$) in a Markov chain. Further, the independence from the chain index $j$ is justified because all chains are probabilistically equivalent. Notably, the covariance at zero lag, $R_{\mathbb{S}_i} (0)$ is equal to $P_{f_i,h}(1-P_{f_i,h})$ since it equals the variance of the failure indicator function (a Bernoulli random variable) for the samples in $\mathbb{S}_i$. Since not all Markov chain states of any $j$th chain necessarily lie in $\mathbb{S}_i$, and more specifically in $\mathcal{W}_i$, let $\pi_i$ denote the set of chain indices with at least one sample in $\mathcal{W}_i$, and $\pi_{ij}$ contain the set of Markov state indices for every $j \in \pi_i$. Then for $2\leq i \leq m$:
\begin{equation}
\begin{split}
\label{sig_i}
        \mathbb{V}(\tilde{P}_{f_i,h}) &= \vartheta_{i,h}^2 = \mathbb{E}\left[ \left(\frac{1}{n_i} \sum_{j \in \pi_i}{\sum_{k \in \pi_{ij}} ({\mathbbm{1}_{fh,jk}^{(i-1)} - P_{f_i,h}})} \right)^2 \right] \\& =
    \frac{1}{n_i^2} \sum_{j\in\pi_i} \mathbb{E}\left[ \left( \sum_{k\in\pi_{ij}}  ({\mathbbm{1}_{fh,jk}^{(i-1)} - P_{f_i,h}})     \right)^2 \right] \\& =
     \frac{1}{n_i^2} \sum_{j\in\pi_i} \sum_{k,k' \in \pi_{ij}} R_{\mathbb{S}_i} (k-k') \\& = 
      \frac{1}{n_i^2} \sum_{j\in\pi_i} R_{\mathbb{S}_i} (0) \psi_{ij} = \frac{P_{f_i,h} (1-P_{f_i,h})}{n_i} \psi_i
\end{split}
\end{equation}
where $\psi_{ij}$ is a linear combination of the ratios ${R_{\mathbb{S}_i} (l)}/{R_{\mathbb{S}_i} (0)}$ whose expression (i.e., the indices $l$ to be evaluated and the corresponding coefficients) depends on $\pi_i, \pi_{ij}$. The intra-stratum correlation is captured by $\psi_i = \sum_{j\in\pi_i}\psi_{ij}/n_i$ based on the intra-chain correlation between the states of the stationary Markov chains. It is clear that the estimator $\tilde{P}_{f_i,h}$ is consistent and that trivially, for the first stratum $\mathbb{V}(\tilde{P}_{f_1,h}) = {P_{f_1,h} (1-P_{f_1,h})}/{n_1}$ which is the MC variance expression. Since inter-chain sample correlation is assumed to be zero, it follows that $\tilde{P}_{f_i,h}$ and $\tilde{P}_{f_j,h}$ are independent. Notably, $\vartheta_{i,h}^2$ accounts for the variability in the MC realizations of the uncertainties in $\pmb{\uptau}$ as well since they are included in the $\pmb{\uptheta}$ samples used to evaluate $\mathbbm{1}_{fh,jk}^{(i-1)}$ while calculating $R_{\mathbb{S}_i}$ of Eq. \eqref{RSi_eqn}.

\subsection{Properties of $\tilde{P}(\mathbb{S}_i)$}

In general, $\tilde{P}(\mathbb{S}_i)$ is asymptotically unbiased as shown below: 
\begin{equation}
    \begin{split}
        \mathbb{E} (\tilde{P}(\mathbb{S}_i)) &= \mathbb{E} (\tilde{P}(F_{i-1})\tilde{P}(\bar{F}_i | F_{i-1}) ) \\& =
        [P(F_{i-1}) + O(1/N)]P(\bar{F}_i | F_{i-1}) \\& = 
        P(\mathbb{S}_i) + O(1/N), \quad 2 \leq i \leq m-1
    \end{split}
\end{equation}
where the overbar denotes the complement of an event, the above properties of $\tilde{P}(F_i | F_{i-1})$ and $\tilde{P}(\bar{F}_{i-1})$ can be noted from the the original subset simulation paper \citep{au2001originalsubset}. Obviously, $\tilde{P}(\mathbb{S}_1)$ is estimated only using MC samples and is unbiased. Also, $\mathbb{E} (\tilde{P}(\mathbb{S}_m)) = \mathbb{E} (\tilde{P}(F_{m-1})) = P(\mathbb{S}_m) + O(1/N)$. Next, expressions for $\mathbb{V}(\tilde{P}(\mathbb{S}_i))$ and the covariance, $\text{Cov}(\tilde{P}(\mathbb{S}_i),\tilde{P}(\mathbb{S}_j))$ are derived in terms of the quantities used in \cite{au2001originalsubset} for the c.o.v of $\tilde{P}(F_i | F_{i-1})$ that are given by:
\begin{equation}
\label{del_i}
    \begin{split}
         \delta_i = \sqrt{\frac{(1-P(F_i | F_{i-1}))(1+\gamma_i)} {NP(F_i | F_{i-1})}}, \quad 1 \leq i \leq m-1
    \end{split}
\end{equation}
where $\gamma_i$ is a correlation factor associated with the samples of $F_{i-1}$ also lying in $F_i$ \citep{au2001originalsubset}. Obviously, $\gamma_1 = 0$. In the following discussion, it is assumed that $\{\tilde{P}(F_i | F_{i-1}),\tilde{P}(F_j | F_{j-1})\}$ are independent for $i \neq j$ which is a reasonable assumption according to \cite{au2001originalsubset}. This also implies that $\tilde{P}(\mathbb{S}_i)$ is unbiased (i.e., eliminating the $O(1/N)$ term). Since $\tilde{P}(\mathbb{S}_i) = \tilde{P}(F_{i-1}) - \tilde{P}(F_i)$, the following can be written for $2 \leq i \leq m-1$:
\begin{equation}
\label{Si_var}
    \begin{split}
         \mathbb{V}(\tilde{P}(\mathbb{S}_i)) &= \vartheta_{\mathbb{S}_i}^2 = \mathbb{V}(\tilde{P}(F_{i-1})) + \mathbb{V}(\tilde{P}(F_{i})) \\&-2\text{Cov}(\tilde{P}(F_{i-1}),\tilde{P}(F_{i-1})\tilde{P}(F_i | F_{i-1})) \\& = 
         P^2(F_{i-1}) (1-2P(F_i|F_{i-1})) \sum_{k=1}^{i-1}\delta_k^2 + P^2(F_i) \sum_{k=1}^i \delta_k^2
    \end{split}
\end{equation}
At the boundaries, $\mathbb{V}(\tilde{P}(\mathbb{S}_1)) = \mathbb{V}(\tilde{P}(F_1)) = \tilde{P}(F_1)(1-\tilde{P}(F_1))/N$ and similarly, $\mathbb{V}(\tilde{P}(\mathbb{S}_m)) = \mathbb{V}(\tilde{P}(F_{m-1})) = P^2(F_{m-1}) \sum_{k=1}^{m-1} \delta_k^2$. Similarly, the covariance $\vartheta_{\mathbb{S}_{ij}}^2 = \text{Cov}(\tilde{P}(\mathbb{S}_i),\tilde{P}(\mathbb{S}_j))$ can be derived as:
\begin{equation}
\label{Si_covar}
    \begin{split}
         \vartheta_{\mathbb{S}_{ij}}^2 &= \mathbb{E}\left(\tilde{P}(\mathbb{S}_i)\tilde{P}(\mathbb{S}_j)\right) - P(\mathbb{S}_i)P(\mathbb{S}_j) \\& = 
         \mathbb{E}\left(\tilde{P}(F_{i-1})\tilde{P}(\bar{F}_i | F_{i-1})\tilde{P}(F_{j-1})\tilde{P}(\bar{F}_j | F_{j-1})\right) - P(\mathbb{S}_i)P(\mathbb{S}_j) \\& = 
         \left(P(F_i|F_{i-1}) - P^2(F_i|F_{i-1})(\delta_i^2 +1)\right) P(\bar{F}_j | F_{j-1}) \\& \times P^2(F_{i-1}) \left(\sum_{k=1}^{i-1}\delta_{k}^2+1\right) \prod_{k=i+1}^{j-1} {P(F_k|F_{k-1})} - P(\mathbb{S}_i)P(\mathbb{S}_j)
    \end{split}
\end{equation}
for the case $1<i<m-1$ and $i<j<m$. For convenience of notation, $\xi_{ij}$ will be defined in the following as:
\begin{equation}
\label{xi_sicovar}
    \xi_{ij} = \left(P(F_i|F_{i-1}) - P^2(F_i|F_{i-1})(\delta_i^2 +1)\right)\prod_{k=i+1}^{j-1} {P(F_k|F_{k-1})}
\end{equation}
For all the possible cases of $i\leq j$, $\left(\vartheta_{\mathbb{S}_{ij}}^2 + P(\mathbb{S}_i)P(\mathbb{S}_j)\right)$ can be written as:
\begin{equation}
\label{Si_covar2}
    \begin{cases}
    \xi_{ij} P^2(F_{i-1}) \left(\sum_{k=1}^{i-1}\delta_{k}^2+1\right)P(\bar{F}_j | F_{j-1}), & 1<i<m-1, \;\;\\ & i<j<m \\
    \xi_{ij} P^2(F_{i-1}) \left(\sum_{k=1}^{i-1}\delta_{k}^2+1\right), & 1<i<m, \;\;\\ & j=m \\
    \xi_{ij} P(\bar{F}_j | F_{j-1}), & i=1, 1<j<m \\
    \xi_{ij}, & i=1, j=m \\
    \vartheta_{\mathbb{S}_i}^2 + P(\mathbb{S}_i)P(\mathbb{S}_j), & i=j
  \end{cases}
\end{equation}
Obviously, the full covariance matrix (i.e., both $i\leq j$ and $i>j$) can be constructed using Eq. \eqref{Si_covar2}.

\subsection{Properties of $\tilde{P}_{f,h}$}    
\label{sec:A3}

It can be shown that the overall estimator is asymptotically unbiased as follows:
\begin{equation}
\label{expexp}
    \begin{split}
        \mathbb{E}({\tilde{P}_{f,h}}) &= \sum_{i=1}^m \mathbb{E}\left( \mathbb{E}\left( {\tilde{P}_{f_i,h} \tilde{P}(\mathbb{S}_i) | \tilde{P}(\mathbb{S}_i)} \right)\right)
        \\& = \sum_{i=1}^m \mathbb{E}\left(P_{f_i,h} \tilde{P}(\mathbb{S}_i)\right) = \sum_{i=1}^m { P_{f_i,h} (P(\mathbb{S}_i) + O(1/N))} \\& = 
        P_{f,h} + O(1/N)
    \end{split}
\end{equation}
While Eq. \eqref{expexp} is generally true, under the additional assumption of independence between $\tilde{P}(F_i | F_{i-1})$ and $\tilde{P}(F_j | F_{j-1}), i\neq j$, the overall estimator is unbiased. The variance of the overall estimator can be decomposed according to the total variance theorem as:
\begin{equation}
\label{finalvareqn}
    \begin{split}
        \mathbb{V}\left(\sum_{i=1}^m \tilde{P}_{f_i,h} \tilde{P}(\mathbb{S}_i)\right) &= \mathbb{E}\left(\mathbb{V}\left(\sum_{i=1}^m \tilde{P}_{f_i,h} \tilde{P}(\mathbb{S}_i) | \tilde{P}(\mathbb{S}_i) \right) \right) \\&+
        \mathbb{V}\left(\mathbb{E}\left(\sum_{i=1}^m \tilde{P}_{f_i,h} \tilde{P}(\mathbb{S}_i) | \tilde{P}(\mathbb{S}_i) \right) \right) \\&=
        \mathbb{E} \left(\sum_{i=1}^m \vartheta_{i,h}^2 \tilde{P}^2(\mathbb{S}_i) \right) +  \mathbb{V}\left(\sum_{i=1}^m P_{f_i,h} \tilde{P}(\mathbb{S}_i) \right) \\&=
        \sum_{i=1}^m \vartheta_{i,h}^2 \left(\vartheta_{\mathbb{S}_i}^2 + P^2(\mathbb{S}_i) \right) \\&+
        \sum_{i=1}^m \sum_{j=1}^m P_{f_i,h}P_{f_j,h} \vartheta_{\mathbb{S}_{ij}}^2
    \end{split}
\end{equation}
Since $\delta_i^2 = O(1/N), \vartheta_{\mathbb{S}_i}^2 = O(1/N)$, and $\vartheta_{i,h}^2 = O(1/n_i)$, it can be seen that $\tilde{P}_{f,h}$ is consistent (i.e., guarantees convergence to true probability as $N \to \infty$ and $n_i \to \infty$).


{\small
\bibliography{Bibliography}

\begin{thebibliography}{}

\bibitem[\protect\citeauthoryear{}{Amelin}{2004}]{amelin2004KTH}
Amelin, M. (2004).
\newblock ``On monte carlo simulation and analysis of electricity markets.''\
  Ph.D. thesis, {KTH Royal Institute of Technology}, Stockholm, Sweden.

\bibitem[\protect\citeauthoryear{}{{American Society of Civil
  Engineers}}{2019}]{Prestandard19}
{American Society of Civil Engineers} (2019).
\newblock {\em Prestandard for Performance-Based Wind Design}.
\newblock Reston, VA.

\bibitem[\protect\citeauthoryear{}{Arnab}{2017}]{arnab2017surveybook}
Arnab, R. (2017).
\newblock {\em Survey sampling theory and applications}.
\newblock Academic Press.

\bibitem[\protect\citeauthoryear{}{Arunachalam and Spence}{2021}]{esrel21}
Arunachalam, S. and Spence, S. M.~J. (2021).
\newblock ``A stochastic simulation scheme for the estimation of small failure
  probabilities in wind engineering applications.''\ {\em Proc. of European
  Safety and Reliability Conference (ESREL 2021)}.

\bibitem[\protect\citeauthoryear{}{{ASCE 7-22}}{2022}]{ASCE7}
{ASCE 7-22} (2022).
\newblock {\em Minimum Design Loads for Buildings and Other Structures}.
\newblock American Society of Civil Engineers (ASCE), Reston, VA.

\bibitem[\protect\citeauthoryear{}{Atkinson and
  Silva}{2000}]{atkinson2000stochastic}
Atkinson, G.~M. and Silva, W. (2000).
\newblock ``{Stochastic modeling of California ground motions}.''\ {\em
  {Bulletin of the Seismological Society of America}}, 90(2), 255--274.

\bibitem[\protect\citeauthoryear{}{Au}{2007}]{au2007augmenting}
Au, S. (2007).
\newblock ``Augmenting approximate solutions for consistent reliability
  analysis.''\ {\em Probabilistic Engineering Mechanics}, 22(1), 77--87.

\bibitem[\protect\citeauthoryear{}{Au and Beck}{2003a}]{au2003importance}
Au, S.-K. and Beck, J. (2003a).
\newblock ``Important sampling in high dimensions.''\ {\em Structural safety},
  25(2), 139--163.

\bibitem[\protect\citeauthoryear{}{Au and Beck}{1999}]{au1999adaptive}
Au, S.-K. and Beck, J.~L. (1999).
\newblock ``A new adaptive importance sampling scheme for reliability
  calculations.''\ {\em Structural safety}, 21(2), 135--158.

\bibitem[\protect\citeauthoryear{}{Au and Beck}{2001}]{au2001originalsubset}
Au, S.-K. and Beck, J.~L. (2001).
\newblock ``Estimation of small failure probabilities in high dimensions by
  subset simulation.''\ {\em Probabilistic engineering mechanics}, 16(4),
  263--277.

\bibitem[\protect\citeauthoryear{}{Au and Beck}{2003b}]{au2003subsetseismic}
Au, S.-K. and Beck, J.~L. (2003b).
\newblock ``Subset simulation and its application to seismic risk based on
  dynamic analysis.''\ {\em Journal of Engineering Mechanics}, 129(8),
  901--917.

\bibitem[\protect\citeauthoryear{}{Baker}{2015}]{baker2015}
Baker, J.~W. (2015).
\newblock ``Efficient analytical fragility function fitting using dynamic
  structural analysis.''\ {\em Earthquake Spectra}, 31(1), 579--599.

\bibitem[\protect\citeauthoryear{}{Bect et~al.\@}{2017}]{bect2017BSS}
Bect, J., Li, L., and Vazquez, E. (2017).
\newblock ``Bayesian subset simulation.''\ {\em SIAM/ASA Journal on Uncertainty
  Quantification}, 5(1), 762--786.

\bibitem[\protect\citeauthoryear{}{Boj{\'o}rquez and
  Ruiz-Garc{\'\i}a}{2013}]{bojorquez2013residual}
Boj{\'o}rquez, E. and Ruiz-Garc{\'\i}a, J. (2013).
\newblock ``Residual drift demands in moment-resisting steel frames subjected
  to narrow-band earthquake ground motions.''\ {\em Earthquake engineering \&
  structural dynamics}, 42(11), 1583--1598.

\bibitem[\protect\citeauthoryear{}{Boore}{2003}]{boore2003simulation}
Boore, D.~M. (2003).
\newblock ``Simulation of ground motion using the stochastic method.''\ {\em
  Pure and applied geophysics}, 160(3), 635--676.

\bibitem[\protect\citeauthoryear{}{Boore and
  Joyner}{1997}]{boorejoyner1997site}
Boore, D.~M. and Joyner, W.~B. (1997).
\newblock ``Site amplifications for generic rock sites.''\ {\em {Bulletin of
  the Seismological Society of America}}, 87(2), 327--341.

\bibitem[\protect\citeauthoryear{}{Chen and Kareem}{2005}]{ChenPOD2005}
Chen, X. and Kareem, A. (2005).
\newblock ``Proper orthogonal decomposition-based modeling, analysis, and
  simulation of dynamic wind load effects on structures.''\ {\em Journal of
  Engineering Mechanics}, 131(4), 325--339.

\bibitem[\protect\citeauthoryear{}{Cochran}{2007}]{cochran2007sampling}
Cochran, W.~G. (2007).
\newblock {\em Sampling techniques}.
\newblock John Wiley \& Sons.

\bibitem[\protect\citeauthoryear{}{Der~Kiureghian}{2022}]{derkiureghian_2022}
Der~Kiureghian, A. (2022).
\newblock {\em {Structural and System Reliability}}.
\newblock Cambridge University Press.

\bibitem[\protect\citeauthoryear{}{Der~Kiureghian and
  Ditlevsen}{2009}]{der2009aleatory}
Der~Kiureghian, A. and Ditlevsen, O. (2009).
\newblock ``{Aleatory or epistemic? Does it matter?}.''\ {\em Structural
  safety}, 31(2), 105--112.

\bibitem[\protect\citeauthoryear{}{Elkady}{2016}]{Elkadythesis}
Elkady, A. (2016).
\newblock ``Collapse risk assessment of steel moment resisting frames designed
  with deep wide-flange columns in seismic regions.''\ Ph.D. thesis, Dept. of
  Civil Engineering and Applied Mechanics, McGill University, Montreal, Canada.

\bibitem[\protect\citeauthoryear{}{Elkady}{2019}]{elkadygithub}
Elkady, A. (2019).
\newblock ``Two-dimensional opensees numerical models for archetype steel
  buildings with special moment frames.''\ {\em GitHub repository}.

\bibitem[\protect\citeauthoryear{}{Elkady and Lignos}{2015}]{elkady2015effect}
Elkady, A. and Lignos, D.~G. (2015).
\newblock ``Effect of gravity framing on the overstrength and collapse capacity
  of steel frame buildings with perimeter special moment frames.''\ {\em
  Earthquake Engineering \& Structural Dynamics}, 44(8), 1289--1307.

\bibitem[\protect\citeauthoryear{}{Evans}{1951}]{evans1951stratification}
Evans, W.~D. (1951).
\newblock ``On stratification and optimum allocations.''\ {\em Journal of the
  American Statistical Association}, 46(253), 95--104.

\bibitem[\protect\citeauthoryear{}{Fishman}{2013}]{fishman2013monte}
Fishman, G. (2013).
\newblock {\em Monte Carlo: concepts, algorithms, and applications}.
\newblock Springer Science \& Business Media.

\bibitem[\protect\citeauthoryear{}{Glasgow}{2005}]{GLASGOW2005}
Glasgow, G. (2005).
\newblock ``Stratified sampling types.''\ {\em Encyclopedia of Social
  Measurement}, Elsevier, New York,  683--688.

\bibitem[\protect\citeauthoryear{}{Gurley and
  Kareem}{1999}]{gurley1999applications}
Gurley, K. and Kareem, A. (1999).
\newblock ``Applications of wavelet transforms in earthquake, wind and ocean
  engineering.''\ {\em Engineering structures}, 21(2), 149--167.

\bibitem[\protect\citeauthoryear{}{Gurley et~al.\@}{1997}]{gurley1997analysis}
Gurley, K.~R., Tognarelli, M.~A., and Kareem, A. (1997).
\newblock ``Analysis and simulation tools for wind engineering.''\ {\em
  Probabilistic Engineering Mechanics}, 12(1), 9--31.

\bibitem[\protect\citeauthoryear{}{Hsu and Ching}{2010}]{hsu2010PSubSim}
Hsu, W.-C. and Ching, J. (2010).
\newblock ``Evaluating small failure probabilities of multiple limit states by
  parallel subset simulation.''\ {\em Probabilistic Engineering Mechanics},
  25(3), 291--304.

\bibitem[\protect\citeauthoryear{}{Iwata
  et~al.\@}{2006}]{iwata2006reparability}
Iwata, Y., Sugimoto, H., and Kuwamura, H. (2006).
\newblock ``Reparability limit of steel structural buildings based on the
  actual data of the hyogoken-nanbu earthquake.''\ {\em Proceedings of the 38th
  Joint Panel. Wind and Seismic effects. NIST Special Publication}, 1057,
  23--32.

\bibitem[\protect\citeauthoryear{}{Jakobsen and
  Madsen}{2004}]{jakobsen2004comparison}
Jakobsen, F. and Madsen, H. (2004).
\newblock ``Comparison and further development of parametric tropical cyclone
  models for storm surge modelling.''\ {\em Journal of Wind Engineering and
  Industrial Aerodynamics}, 92(5), 375--391.

\bibitem[\protect\citeauthoryear{}{Jalayer and
  Beck}{2006}]{jalayer2006information}
Jalayer, F. and Beck, J. (2006).
\newblock ``Using information theory concepts to compare alternative intensity
  measures for representing ground motion uncertainty.''\ {\em Proc., 8th US
  National Conf. Earthquake Engineering, Paper ID 974}.

\bibitem[\protect\citeauthoryear{}{Koutsourelakis
  et~al.\@}{2004}]{koutsourelakis2004linesampling}
Koutsourelakis, P.-S., Pradlwarter, H.~J., and Schueller, G.~I. (2004).
\newblock ``{Reliability of structures in high dimensions, part I: algorithms
  and applications}.''\ {\em Probabilistic Engineering Mechanics}, 19(4),
  409--417.

\bibitem[\protect\citeauthoryear{}{Kramer}{2003}]{kramer_2003}
Kramer, S.~L. (2003).
\newblock {\em Geotechnical earthquake engineering}.
\newblock Prentice Hall.

\bibitem[\protect\citeauthoryear{}{Li}{2022}]{li2022rapid}
Li, B. (2022).
\newblock ``Rapid stochastic response estimation of dynamic nonlinear
  structures: Innovative frameworks and applications.''\ Ph.D. thesis,
  {University of Michigan}, Ann Arbor, MI.

\bibitem[\protect\citeauthoryear{}{Li et~al.\@}{2021}]{li2021adaptive}
Li, B., Chuang, W.-C., and Spence, S.~M. (2021).
\newblock ``{An adaptive fast nonlinear analysis (AFNA) algorithm for rapid
  time history analysis}.''\ {\em 8th ECCOMAS Thematic Conference on
  Computational Methods in Structural Dynamics and Earthquake Engineering}.

\bibitem[\protect\citeauthoryear{}{Li et~al.\@}{2017}]{li2017systemGSS2}
Li, D.-Q., Yang, Z.-Y., Cao, Z.-J., Au, S.-K., and Phoon, K.-K. (2017).
\newblock ``System reliability analysis of slope stability using generalized
  subset simulation.''\ {\em Applied Mathematical Modelling}, 46, 650--664.

\bibitem[\protect\citeauthoryear{}{Li et~al.\@}{2015}]{li2015GSS}
Li, H.-S., Ma, Y.-Z., and Cao, Z. (2015).
\newblock ``A generalized subset simulation approach for estimating small
  failure probabilities of multiple stochastic responses.''\ {\em Computers \&
  Structures}, 153, 239--251.

\bibitem[\protect\citeauthoryear{}{Li et~al.\@}{2022}]{li2022explicit}
Li, X.-W., Zhao, Y.-G., Zhang, X.-Y., and Lu, Z.-H. (2022).
\newblock ``Explicit model of outcrossing rate for time-variant reliability.''\
  {\em ASCE-ASME Journal of Risk and Uncertainty in Engineering Systems, Part
  A: Civil Engineering}, 8(1), 04021087.

\bibitem[\protect\citeauthoryear{}{Mazzoni et~al.\@}{2006}]{OpenSees}
Mazzoni, S., McKenna, F., Scott, M.~H., and Fenves, G.~L. (2006).
\newblock {\em OpenSees command language manual}.
\newblock University of California, Berkeley, CA.

\bibitem[\protect\citeauthoryear{}{Melchers}{1989}]{melchers1989importance}
Melchers, R. (1989).
\newblock ``Importance sampling in structural systems.''\ {\em Structural
  safety}, 6(1), 3--10.

\bibitem[\protect\citeauthoryear{}{Melchers and
  Beck}{2018}]{melchers2018structural}
Melchers, R.~E. and Beck, A.~T. (2018).
\newblock {\em Structural reliability analysis and prediction}.
\newblock John wiley \& sons.

\bibitem[\protect\citeauthoryear{}{Neyman}{1934}]{neyman1934}
Neyman, J. (1934).
\newblock ``On the two different aspects of the representative method: The
  method of stratified sampling and the method of purposive selection.''\ {\em
  Journal of the Royal Statistical Society}, 97(4), 558--606.

\bibitem[\protect\citeauthoryear{}{Ouyang and
  Spence}{2021}]{ouyang2021performance}
Ouyang, Z. and Spence, S.~M. (2021).
\newblock ``A performance-based wind engineering framework for engineered
  building systems subject to hurricanes.''\ {\em Frontiers in Built
  Environment},  133.

\bibitem[\protect\citeauthoryear{}{Papaioannou
  et~al.\@}{2015}]{papaioannou2015mcmc}
Papaioannou, I., Betz, W., Zwirglmaier, K., and Straub, D. (2015).
\newblock ``{MCMC algorithms for subset simulation}.''\ {\em Probabilistic
  Engineering Mechanics}, 41, 89--103.

\bibitem[\protect\citeauthoryear{}{Papaioannou
  et~al.\@}{2016}]{papaioannou2016sequentialIS}
Papaioannou, I., Papadimitriou, C., and Straub, D. (2016).
\newblock ``Sequential importance sampling for structural reliability
  analysis.''\ {\em Structural safety}, 62, 66--75.

\bibitem[\protect\citeauthoryear{}{Pharr et~al.\@}{2017}]{PHARR2017747}
Pharr, M., Jakob, W., and Humphreys, G. (2017).
\newblock ``{13 - Monte Carlo Integration}.''\ {\em Physically Based
  Rendering}, M. Pharr, W. Jakob, and G. Humphreys, eds., Morgan Kaufmann,
  Boston, third edition edition,  747--802.

\bibitem[\protect\citeauthoryear{}{Rao}{1973}]{Rao1973double}
Rao, J. N.~K. (1973).
\newblock ``On double sampling for stratification and analytical surveys.''\
  {\em Biometrika}, 60(1), 125--133.

\bibitem[\protect\citeauthoryear{}{Rice}{1944}]{rice1944mathematical}
Rice, S.~O. (1944).
\newblock ``Mathematical analysis of random noise.''\ {\em The Bell System
  Technical Journal}, 23(3), 282--332.

\bibitem[\protect\citeauthoryear{}{Schueller
  et~al.\@}{2004}]{schueller2004critical}
Schueller, G.~I., Pradlwarter, H.~J., and Koutsourelakis, P.-S. (2004).
\newblock ``A critical appraisal of reliability estimation procedures for high
  dimensions.''\ {\em Probabilistic engineering mechanics}, 19(4), 463--474.

\bibitem[\protect\citeauthoryear{}{Shields and Zhang}{2016}]{shields2016}
Shields, M.~D. and Zhang, J. (2016).
\newblock ``{The generalization of Latin hypercube sampling}.''\ {\em
  Reliability Engineering \& System Safety}, 148, 96--108.

\bibitem[\protect\citeauthoryear{}{Shinozuka and
  Deodatis}{1991}]{shinozuka1991simulation}
Shinozuka, M. and Deodatis, G. (1991).
\newblock ``Simulation of stochastic processes by spectral representation.''\
  {\em Applied Mechanics Reviews}, 44(4), 191--204.

\bibitem[\protect\citeauthoryear{}{Shinozuka and
  Deodatis}{1996}]{shinozuka1996fields}
Shinozuka, M. and Deodatis, G. (1996).
\newblock ``Simulation of multi-dimensional gaussian stochastic fields by
  spectral representation.''\ {\em Applied Mechanics Reviews}, 49(1), 29--53.

\bibitem[\protect\citeauthoryear{}{Shome et~al.\@}{1998}]{shome1998earthquakes}
Shome, N., Cornell, C.~A., Bazzurro, P., and Carballo, J.~E. (1998).
\newblock ``Earthquakes, records, and nonlinear responses.''\ {\em Earthquake
  spectra}, 14(3), 469--500.

\bibitem[\protect\citeauthoryear{}{Stein}{1987}]{stein1987large}
Stein, M. (1987).
\newblock ``{Large sample properties of simulations using Latin hypercube
  sampling}.''\ {\em Technometrics}, 29(2), 143--151.

\bibitem[\protect\citeauthoryear{}{Sudret}{2012}]{sudretmetamodel}
Sudret, B. (2012).
\newblock ``Meta-models for structural reliability and uncertainty
  quantification.''\ {\em arXiv preprint arXiv:1203.2062}.

\bibitem[\protect\citeauthoryear{}{Vetter and
  Taflanidis}{2012}]{vetter2012global}
Vetter, C. and Taflanidis, A.~A. (2012).
\newblock ``Global sensitivity analysis for stochastic ground motion modeling
  in seismic-risk assessment.''\ {\em Soil Dynamics and Earthquake
  Engineering}, 38, 128--143.

\bibitem[\protect\citeauthoryear{}{Vickery
  et~al.\@}{2000}]{vickery2000simulation}
Vickery, P., Skerlj, P., and Twisdale, L. (2000).
\newblock ``{Simulation of hurricane risk in the US using empirical track
  model}.''\ {\em Journal of structural engineering}, 126(10), 1222--1237.

\bibitem[\protect\citeauthoryear{}{Vickery and
  Twisdale}{1995a}]{vickery1995prediction}
Vickery, P.~J. and Twisdale, L.~A. (1995a).
\newblock ``{Prediction of hurricane wind speeds in the United States}.''\ {\em
  Journal of Structural Engineering}, 121(11), 1691--1699.

\bibitem[\protect\citeauthoryear{}{Vickery and
  Twisdale}{1995b}]{vickery1995wind}
Vickery, P.~J. and Twisdale, L.~A. (1995b).
\newblock ``Wind-field and filling models for hurricane wind-speed
  predictions.''\ {\em Journal of Structural Engineering}, 121(11), 1700--1709.

\end{thebibliography}
}
\end{document}